\documentclass[a4paper,12pt,titlepage]{article}
\usepackage[utf8]{inputenc}
\usepackage{amsmath}
\usepackage{amsfonts}
\usepackage{amssymb}
\usepackage{graphicx}
\usepackage{setspace}
\usepackage{adjustbox}
\usepackage{xcolor}
\usepackage{csquotes}
\usepackage{listings}
\usepackage[backend=bibtex, style=authoryear, maxcitenames=3, maxbibnames=6, minbibnames=6, giveninits=true, uniquename=init]{biblatex}
\RequirePackage[rmargin=1in,lmargin=1in, lines=25]{geometry}
\usepackage{authblk}

\newcommand*\oline[1]{%
   \vbox{%
     \hrule height 0.03pt
     \kern0.35ex
     \hbox{%
       \kern-0.2em
       \ifmmode#1\else\ensuremath{#1}\fi
       \kern-0.2em
     }
   }
}

\title{Bayesian semi-parametric G-computation for causal inference in a cohort study with MNAR dropout and death.}

\author[1,*]{Josefsson, Maria}

\author[2]{Daniels, Michael J.}

\affil[1]{Centre for Demographic and Ageing Research,\protect\\
Ume{\aa} University, Sweden.}

\affil[2]{Department of Statistics, University of Florida.}

\affil[*]{Corresponding author: Maria Josefsson,\protect\\
Umeå University, SE-901 87 Umeå, Sweden.\protect\\
E-mail: maria.josefsson@umu.se}
\date{}

\begin{document}

\maketitle

\newpage
\section*{Abstract} 
Causal inference with observational longitudinal data and time-varying exposures is often complicated by time-dependent confounding and attrition. The G-computation formula is one approach for estimating a causal effect in this setting. The parametric modeling approach typically used in practice relies on strong modeling assumptions for valid inference, and moreover depends on an assumption of missing at random, which is not appropriate when the missingness is missing not at random (MNAR) or due to death. In this work we develop a flexible Bayesian semi-parametric G-computation approach for assessing the causal effect on the subpopulation that would survive irrespective of exposure, in a setting with MNAR dropout. The approach is to specify models for the observed data using Bayesian additive regression trees, and then use assumptions with embedded sensitivity parameters to identify and estimate the causal effect. The proposed approach is motivated by a longitudinal cohort study on cognition, health, and aging, and we apply our approach to study the effect of becoming a widow on memory. We also compare our approach to several standard methods.

\subsection*{Keywords} 
BART, Cognitive aging, Longitudinal data, Observational data, Non-ignorable missing, Sensitivity analysis, Survivor Average Causal Effect, Time-varying exposure, Time-varying confounding.
\newpage

\section{Introduction} 
Causal inference in non-randomized longitudinal studies with time-varying exposures is often complicated by time-dependent confounding and attrition. Attrition is inevitable especially if individuals in the studied population are older and followed over a long time period. Additionally, for cohort studies, an individual's data is only recorded if that person completes follow-up testing. Hence, data for not only the outcome but also exposure level and confounders are missing at subsequent test waves. 

The G-computation formula (\cite{robins1986new}) is one approach for estimating a causal effect of time-varying exposures when time-varying confounding is present. The approach is completely nonparametric in its original form, although a parametric modeling approach based on maximum likelihood estimation is most typically used in practice  (e.g. \cite{snowden2011implementation, wang2017g}). Valid inference with the parametric G-formula requires correct model specification. This can be extremely difficult when there is a large set of regressors, the relationship is non-linear and/or includes interaction terms, and there are multiple observation times. 
Non- and semi-parametric estimation techniques that do not require prespecified distributional or functional forms of the data, have become popular in the causal inference literature (e.g. \cite{hill2011bayesian, haggstrom2018data, kim2017framework, karim2017application, tan2019BART, wager2018estimation}). One such modeling strategy is Bayesian Additive Regression Trees (BART, \cite{chipman2010bart}). BART is a sum-of-trees model that adds together the predictions of a number of regression trees regularized by prior distributions. BART does not rely on strong modeling assumptions, and in contrast to other tree-based algorithms BART yields interval estimates for full posterior inference.

A number of methodologies have been applied to address missing response or missing covariate data in causal effect estimation of longitudinal data under an assumption of missing at random (MAR; \cite{chen2011doubly, robins1995analysis}). These methods, however, are generally invalid when the missingness is missing not at random (MNAR) or due to death (\cite{kurland2009longitudinal}).
Partly conditional models have been proposed to address the combination of dropout and truncation by death, where inference is conditioning on the sub-population being alive at a specific time-point (\cite{kurland2005directly, shardell2008weighted, li2018accommodating, rizopoulos2012joint, wen2018semi})
However, conditioning on survival may introduce bias due to the fact that survival is a post-randomization event. One estimand that has gained much attention to address this issue is the "survivors average causal effect" (SACE), i.e. the causal effect on the subpopulation of those surviving irrespective of exposure (\cite{frangakis02, frangakis07}). Several approaches have been developed for estimation of the SACE in longitudinal randomized control studies (e.g. \cite{lee2013causal, lee2010causal, wang2017inference, wang2017causal}), or in context of semicompeting risks (\cite{comment2019survivor, xu2019bayesian}). For observational data \citeauthor{tchetgen2014identification} (\citeyear{tchetgen2014identification}) developed a weighting estimator to identify the SACE without missingness, and \citeauthor{shardell2014doubly} (\citeyear{shardell2014doubly}) identified the SACE with MAR missingness using also a weighting technique. Moreover, \citeauthor{josefsson2016causal} (\citeyear{josefsson2016causal}) proposed assumptions to identify the SACE of a baseline exposure on a longitudinal outcome under MNAR missingness for the outcome using parametric methods. These approaches however, do not appropriately account for MNAR data among survivors when the exposure and confounding are time-varying. \citeauthor{shardell2018joint} (\citeyear{shardell2018joint}) proposed a parametric shared parameter model with g-computation to identify a principal stratum causal effect for observational longitudinal data with time-dependent confounding. A drawback of their approach is that unbiased estimation depends on correct model specification and it does not appropriately account for MNAR data among survivors. 

Widowhood has been identified as an important social factor associated with increased mortality (\cite{haakansson2009association}) and cognitive impairment (e.g. \cite{mousavi12}). Here, our goal is to develop a framework for assessing the impact of becoming a widow on memory, a monotone exposure, by estimating the SACE in a setting with MNAR dropout among survivors.  The proposed approach is motivated by the Betula study (\cite{nilsson97}), where individuals are followed over multiple test waves to study how cognitive functions potentially deteriorate with age and identify risk factors for dementia. 

The remainder of the paper is organized as follows. In Section 2, we introduce the notation and the causal estimand. In Section 3, we propose identifying default assumptions and sensitivity parameters to allow deviations from these assumptions, followed by the identification of the SACE in Section 4. In Section 5, we propose a Bayesian semi-parametric (BSP) modeling approach for the observed data distributions and the algorithm for estimation of the SACE. In Section 6, we provide a simulation and in Section 7 an application to the Betula data. Finally, we conclude with a discussion and possible future work in Section 8.
\section{Notation and the causal effect of interest}
\subsection{Data structure and notation}
\begin{figure}
\begin{center}
\caption{\label{Dag} A causal diagram of a simplified version of the Betula study design restricted to two test waves. }
\includegraphics[scale=1]{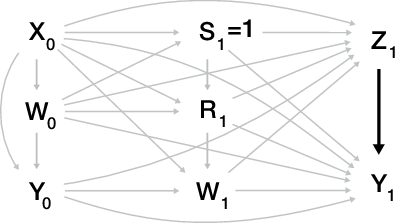} 
\end{center}
\end{figure}

We begin with a formal description of the data. Let $i = 1, 2, \ldots, N$ denote individual and $j=0, 1, \ldots,J$ denote time (the data used from the Betula study has $J=3$ follow-up test waves). We denote the vector of baseline confounders by $X_{i0}$ (gender, education, and age cohort) and the time-varying confounder by $W_{ij}$ (if the spouse has been seriously ill between the $j-1$th and $j$th test wave). The continuous memory outcome is denoted by $Y_{ij}$ and the binary exposure is denoted by $Z_{ij}$.
We assume a monotone exposure where initially all subjects are unexposed ($Z_{10}=0$ for all $i$). If a subject is exposed (widowed) at test wave $j$ $Z_{ij}=1$ and if $Z_{ij}=1$, then $Z_{ik}=1$ for $k > j$.
Let $S_{ij}$ denote survival, where $S_{ij}=1$ if an individual is alive at the time of the testing and 0 otherwise. Let $R_{ij}$ be a dropout indicator, where $R_{ij}=1$ if an individual has completed the cognitive testing or 0 otherwise. We have monotone missingness, so if $R_{ij}=0$, $R_{ik}=0$ for $k > j$. Note that vital status information is presumed to be available even after dropping out of the study.
The history of the time-varying variables are denoted with an overbar. For example, the exposure history for individual $i$ through test wave $j$ is denoted by $\bar{Z}_{ij}=\{Z_{i0},Z_{i1},\ldots,Z_{ij}\}$.  
Furthermore, for individual $i$, $J^r_i$ denotes the number of test waves (s)he participates in the study, and $J^s_i\geq J^r_i$ denotes the number of test waves (s)he is alive. A simplified version of the study design restricted to two test waves is depicted in a causal diagram in Figure \ref{Dag}. 

\subsection{Causal estimand} 
The goal of the study is to estimate the causal effect of becoming a widow (within 5 years) on memory among those who would survive irrespective of being widowed or not. We consider two contrasting exposure regimes, $\bar{z}_{ij}=\{z_{i0}=0,\ldots,z_{ij-1}=0,z_{ij}=1\}$, i.e individuals exposed (widowed) between the $j-1$th and $j$th wave,  and the contrasting regime $\bar{z}'_{ij}=\{z_{i0}=0,\ldots,z_{ij}=0\}$, i.e. individuals unexposed through test wave $j$, for $j=1,2,3$.
Below, we generally suppress the subscript $i$ to simplify notation. The potential memory outcome at wave $j$ is denoted by $Y_j(\bar{z}_j)$ for an individual under exposure regime $\bar{z}_j$. Similarly, let $S_j(\bar{z}_j)$ be the potential survival outcome at wave $j$, denoting survival under exposure regime $\bar{z}_j$. 

We consider a principal stratum causal effect of a time-varying exposure on the outcome, at wave $j$, for those who would survive under either exposure regime, 
\begin{equation}\label{ident.eq}
\mathrm{E}[Y_j(\bar{z}_j)-Y_j(\bar{z}'_j)\mid \bar{S}_j(\bar{z}_j) = \bar{S}_j(\bar{z}'_j)= 1].
\end{equation} 
However, main interest is not the effect at a specific wave, but rather the effect aggregated over test waves, defined as
\begin{equation}\label{Tau.eq}
\tau = \frac{\sum _{j=1}^J \mathrm{E}[Y_j(\bar{z}_j)-Y_j(\bar{z}'_j)\mid \bar{S}_j(\bar{z}_j) = \bar{S}_j(\bar{z}'_j)= 1] \times \Pr[\bar{S}_j(\bar{z}_j) = \bar{S}_j(\bar{z}'_j)= 1]}{\sum_{k=1}^J \Pr[\bar{S}_k(\bar{z}_k) = \bar{S}_k(\bar{z}_k^\prime)= 1]}.
\end{equation} 
  
\section{Identifying assumptions and sensitivity parameters}
To identify the causal effect in [\ref{Tau.eq}] from the observed data we first introduce a set of assumptions followed by a set of sensitivity parameters to assess the impact of violations to some of the assumptions. The sensitivity parameters (and their values) will be explained in relation to the Betula data in Section 7.2.

\subsection{Assumptions}
Assumptions $1-3$ are a set of standard assumptions for causal inference of longitudinal observational data:

\textbf{Assumption 1} \emph{Consistency}: For a given individual, if $\bar{Z}_{j}=\bar{z}_{j}$, then $Y_j=Y_j(\bar{z}_{j})$ and $S_j=S_j(\bar{z}_{j})$.

\textbf{Assumption 2} \emph{Positivity for a monotone exposure}: $\Pr[z_{j} \mid \bar{y}_{j-1}, \bar{z}_{j-1}=0, \bar{w}_{j}, \bar{r}_{j-1}, \bar{s}_{j-1}=1, x_0] > 0$ for $z_j=0,1$ and for all individuals, such that all unexposed individuals have a nonzero probability of becoming exposed between test wave $j-1$ and $j$ if $p(\bar{y}_{j-1}, \bar{z}_{j-1}, \bar{w}_{j}, \bar{r}_{j-1}, \bar{s}_{j-1}=1, x_0)\neq 0$. 

\textbf{Assumption 3} \emph{Conditional exchangeability}: If $X_0$ and $\oline{W}_{j}$ contains all pre-exposure covariates related to exposure, potential outcomes and survival, then for all exposure regimes 
\begin{align*}
&Y_j(\bar{z}_j)  \perp\!\!\!\perp Z_j \mid \bar{y}_{j-1}, \bar{z}_{j-1}, \bar{w}_{j}, \bar{r}_{j}, \bar{s}_j=1, x_0 \\
&S_j(\bar{z}_j)  \perp\!\!\!\perp Z_j \mid \bar{y}_{j-1}, \bar{z}_{j-1}, \bar{w}_{j}, \bar{r}_{j}, \bar{s}_{j-1}=1, x_0.
\end{align*} 
That is, at each test wave $j$, being exposed $z_j$ is as if randomized conditional on the set of the temporally preceding variables. The assumption of conditional exchangeability is likely to be violated in many settings and is impossible to assess from the data. Therefore, we introduce a sensitivity parameter to investigate sensitivity for unmeasured confounding in Section 3.2. 

In cohort studies $Y_j, Z_j$ and $W_j$ are not observed (but defined) for individuals who are alive but who drop out of the study. We make an MAR type assumption conditional on being survival at time $j$ (MAR-S) to identify the distribution of dropouts among survivors. 

\textbf{Assumption 4} \emph{Dropout among survivors}
For all $j\geq1$ and all $t \leq j$
$$p(y_j \mid \bar{y}_{j-1}, \bar{z}_{j},\bar{w}_{j}, r_j=0, \bar{s}_{j}=1, x_0) = p(y_j \mid \bar{y}_{j-1}, \bar{z}_{j}, \bar{w}_{j}, \bar{r}_{j}=1, \bar{s}_j=1, x_0)$$
That is, the outcome is distributed the same among dropouts and non-dropouts conditional on survival and the temporally preceding variables.
Similarly, $p(w_j \mid \bar{y}_{j-1}, \bar{z}_{j-1},\bar{w}_{j-1}, r_j=0,  \bar{s}_{j}=1,x_0) = p(w_j \mid \bar{y}_{j-1}, \bar{z}_{j-1},\bar{w}_{j-1}, \bar{r}_{j}=1, \bar{s}_{j}=1,x_0)$ and $p(z_j \mid \bar{y}_{j-1}, \bar{z}_{j-1},\bar{w}_{j}, r_j=0,  \bar{s}_{j}=1, x_0) =p(z_j \mid \bar{y}_{j-1}, \bar{z}_{j-1}, \bar{w}_{j}, \bar{r}_{j}=1, \bar{s}_j=1, x_0)$. Previous studies of the Betula data have shown that individuals who drop out have lower cognitive performance and steeper decline (\cite{josefsson12}). In Section 3.2 we introduce sensitivity parameters to allow the dropout to deviate from this MAR type assumption.

We also need three further assumptions for identification of the potential outcomes for those individuals who would survive regardless of exposure history, i.e. the principal strata. We start with two standard assumptions.

\textbf{Assumption 5} \emph{Monotonicity}. $S_{j}(\bar{z}_{j}) \leq S_{j}(\bar{z}'_{j})$; if an individual were to be alive under exposure regime $\bar{z}_j$ then (s)he would also be alive under the contrasting regime $\bar{z}'_j$. Deterministic monotonicity can be too strong in many settings and we discuss a weakening of this in Section 8.

\textbf{Assumption 6} \emph{Differences in outcomes when comparing different strata}. For the contrasting exposure regime $\bar{z}'_{j}$ we assume,
$E[Y_j(\bar{z}'_{j}) \mid \bar{S}_j(\bar{z}_j) = \bar{S}_j(\bar{z}'_j)= 1] = E[Y(\bar{z}'_{j}) \mid \bar{S}_j(\bar{z}'_j) = 1, \bar{S}_j(\bar{z}_j) \neq 1]$. That is, there is no difference in potential outcomes when comparing the "always survivor" strata to the strata where individuals were to live under the contrasting regime $\bar{z}'_{j}$ but not under exposure regime $\bar{z}_{j}$.
In Section 3.2 we introduce a sensitivity parameter to investigate sensitivity to this assumption, due to the fact that individuals in the always survivor strata are likely healthier and have better cognitive performance.

A common problem encountered in longitudinal cohort studies is that an individual's exposure level $z_j$, hence the exposure regime $\bar{z}_j$, and time-varying confounder $w_j$ is only observed if (s)he is alive and participates at the $j$th test wave. Hence we need to introduce a new assumption to be able to identify the probability of survival among exposed and non-exposed; this is necessary for the identification of the potential outcomes among always survivors. 

\textbf{Assumption 7} \emph{Exposure and confounding among non-survivors}
If $s_j=0$ and $\bar{s}_{j-1}=1$ for an individual, $z_{j}$ and $w_j$ may have occurred before the event of death, thus, $z_{j}$ and $w_j$ are not observed but could still be well-defined. We assume,
\begin{align*}
&\Pr[z_{j} \mid \bar{y}_{j-1}, \bar{z}_{j-1},\bar{w}_{j},r_j=0,\bar{r}_{j-1}, s_j=0, \bar{s}_{j-1}=1, x_0] \\
&= \Pr[z_{j} \mid \bar{y}_{j-1}, \bar{z}_{j-1},\bar{w}_{j},\bar{r}_{j}, \bar{s}_{j}=1, x_0],
\end{align*}
and
\begin{align*}
&Pr[w_j \mid \bar{y}_{j-1}, \bar{z}_{j-1},\bar{w}_{j-1},r_{j}=0, \bar{r}_{j-1},s_{j}=0, \bar{s}_{j-1}= 1, x_0] \\
&= \Pr[w_j \mid \bar{y}_{j-1},\bar{z}_{j-1},\bar{w}_{j-1},\bar{r}_j,\bar{s}_{j}= 1, x_0],
\end{align*}
i.e. the exposure and confounder are distributed the same among survivors and non-survivors conditional on the temporally preceding variables. 

This assumption is used for identification of the principal strata. In the Betula study the cognitive testing is performed at 5 year intervals. Since 5 years is a rather long time period it is likely that some of the participants who died before follow-up were also widowed before death. Thus, the number of widowed participants in the sample may be underestimated and must be accounted for.

\subsection{Sensitivity parameters}
To investigate sensitivity of Assumption 3 we follow the procedure of \citeauthor{brumback2004sensitivity} (\citeyear{brumback2004sensitivity}). The unmeasured confounding is quantified through a parameter which describes the outcome confounding. That is, for exposure regime $\bar{z}_{j}$,  
$c(\bar{z}_j)= E[ Y_j(\bar{z}_j) \mid \bar{y}_{j-1}, \bar{z}_j, \bar{w}_{j}, \bar{r}_{j}, \bar{s}_j=1, x_0] - E[Y_j(\bar{z}_j) \mid \bar{y}_{j-1}, \bar{z}'_j, \bar{w}_{j}, \bar{r}_{j}, \bar{s}_j=1, x_0]$, where $c(\bar{z}_j)$ is the average difference in potential outcomes because of unmeasured confounding. The conditional exchangeability assumption does not hold if $c(\bar{z}_j)\neq 0$. Thus, estimating $E[ Y_j(\bar{z}_j) \mid \bar{y}_{j-1}, \bar{w}_{j}, \bar{r}_{j}, \bar{s}_j=1, x_0]$ using the naive estimator $E[Y_j\mid \bar{y}_{j-1}, \bar{z}_{j}, \bar{w}_{j}, \bar{r}_{j}, \bar{s}_j=1, x_0]$ leads to a bias of $c(\bar{z}_j)\times \Pr[z'_j \mid \bar{y}_{j-1}, \bar{z}_{j-1}, \bar{w}_{j}, \bar{r}_{j}, \bar{s}_j=1, x_0]$. 
Further, since the two regimes only differ in $z_j$, for $\bar{z}'_j$, the bias becomes $c(\bar{z}'_j)\times \Pr[z_j \mid \bar{y}_{j-1}, \bar{z}'_{j-1}, \bar{w}_{j}, \bar{r}_{j}, \bar{s}_j=1, x_0]
$. Sensitivity to several types of unmeasured confounding can be assessed using this form. Here, we restrict to an unmeasured confounder independent of the history of the joint processes $(\bar{y}_{j-1}, \bar{z}_j, \bar{w}_{j}, \bar{r}_{j}, \bar{s}_j,x_0)$.

To investigate sensitivity of Assumption 4 we first make an assumption of non-future dependence (NFD) conditional on survival (NFD-S) for the outcome and then instroduce sensitivity parameters within this partial identifying restrictions (\cite{linero2018bayesian}). NFD is a special case of MNAR (\cite{kenward2003pattern}), and NFD-S is defined as,
$p(y_j \mid \bar{y}_{j-1}, \bar{z}_{j},\bar{w}_{j}, \{r_0=1,\ldots,r_{t-1}=1,r_t=0,\ldots,r_j=0 \}, \bar{s}_j=1, x_0) =p(y_j \mid \bar{y}_{j-1}, \bar{z}_{j},\bar{w}_{j}, \bar{r}_j=1, \bar{s}_j=1, x_0),$ for all $j>1$ and all $t < j$. Here it is defined conditional on being alive at time $j$. The NFD-S assumption leaves one conditional distribution per incomplete dropout pattern unidentified, that is when $t=j$. To identify the unidentified conditional distribution left by the NFD-S assumption, we introduce a sensitivity parameter $\gamma_j$ such that
$p(y_{j} \mid \bar{y}_{j-1}, \bar{z}_{j},\bar{w}_{j},\bar{r}_j=\{1,\ldots,1,0 \}, \bar{s}_j=1, x_0) = p(y_j + \gamma_j \mid \bar{y}_{j-1}, \bar{z}_{j},\bar{w}_{j},\bar{r}_j=1, \bar{s}_j=1, x_0)$, when $\gamma_j<0$ implies a negative location shift in the outcome at the first unobserved test wave. This assumption implies dropout at time $j$ depends on being alive at that time, the history up to that time, the exposure, time-varying confounder and the outcome at time $j$, but not outcomes or time-varying variables after time $j$. This assumption of dropout not depending on the 'future' is often viewed as realistic and was proposed originally as a remedy to concerns about many pattern mixture models implicitly having future dependence. Table \ref{dropoutmortality.pat} displays a description of the possible mortality- and missing data patterns under the NFD-S assumption. 

To investigate sensitivity of Assumption 6 we let,
$\Delta_{\bar{z}'_{j}} = E[Y_j(\bar{z}'_{j}) \mid \bar{S}_j(\bar{z}_j) = \bar{S}_j(\bar{z}'_j)= 1] - E[Y(\bar{z}'_{j}) \mid \bar{S}_j(\bar{z}'_j) = 1, \bar{S}_j(\bar{z}_j) \neq 1]$, for the contrasting exposure regime $\bar{z}'_{j}$. That is, the mean difference in potential outcomes when comparing the "always survivor" strata to the strata where individuals were to live under the contrasting regime $\bar{z}'_{j}$ but not under exposure regime $\bar{z}_{j}$. In our analysis we assume $\Delta_{\bar{z}'_{j}}\geq 0$ which implies that memory performance is on average higher in the "always survivors"-strata (the always survivors-strata is healthier). We further assume this difference is independent of the preceding variables. 

To investigate sensitivity of Assumption 7, we introduce a sensitivity parameter $\nu_j$ for the exposure such that,
\begin{align*}
\nu_j = & \Pr[z_{j} \mid \bar{y}_{j-1}, \bar{z}_{j-1},\bar{w}_{j},r_j=0,\bar{r}_{j-1}, s_j=0, \bar{s}_{j-1}=1, x_0] - \\
& \Pr[z_{j} \mid \bar{y}_{j-1}, \bar{z}_{j-1},\bar{w}_{j},\bar{r}_{j}, \bar{s}_{j}=1, x_0],
\end{align*}
representing the mean difference in the proportion exposed between non-survivors and survivors. The first probability on the right-hand side of each expression is not identified. However, bounds can be derived for $\nu_j$; see the Web Appendix section A.2 for details. In particular, the upper bound for  $\nu_j$, $U_{\nu_j}$, is obtained when $\Pr[z_{j} \mid \bar{y}_{j-1}, \bar{z}_{j-1},\bar{w}_{j},r_j=0,\bar{r}_{j-1},s_j=0, \bar{s}_{j-1}=1, x_0]=1$. This reflects that among non-survivors, all subjects were exposed before the event of death between the $j-1$th and $j$th wave. Further, by using Assumption 1 and 5, the lower bound for $\nu_j$ is obtained when 
$\Pr[S_j=1 \mid \bar{y}_{j-1}, \bar{z}_{j},\bar{w}_{j},\bar{r}_j,\bar{s}_{j-1}=1, x_0]=\Pr[S_j=1 \mid \bar{y}_{j-1}, \bar{z}'_{j},\bar{w}_{j},\bar{r}_j,\bar{s}_{j-1}=1, x_0]$. This reflects an equal survival probability among those exposed or unexposed at wave $j$. Here, by using the law of total probability and Bayes theorem, the lower bound $L_{\nu_j}$ becomes 0.

\section{Identification}
Identification of the SACE in [\ref{Tau.eq}] follows from two results.
\begin{description}
\item[Result 1:] The causal contrasts in [\ref{ident.eq}] can be identified as follows
\begin{align}\label{sace.eq}
&E[Y_j(\bar{z}_j) - Y_j(\bar{z}'_j) \mid \bar{S}(\bar{z}_j)=\bar{S}_j(\bar{z}'_j)=1]= \nonumber \\ 
& \quad \frac{E_{\mathcal{A}}[E(Y_j, \bar{S}_j=1\mid \bar{z}_j, c(\bar{z}_j), \bar{\gamma_j},\mathcal{A})]}{E_{\mathcal{A}}[\Pr(\bar{S}_j=1\mid \bar{z}_j,c(\bar{z}_j), \bar{\gamma_j},\nu_j,\mathcal{A})]} - \frac{E_{\mathcal{A}}[E(Y_j, \bar{S}_j=1\mid \bar{z}'_j,c(\bar{z}'_j), \bar{\gamma_j},\mathcal{A})]}{E_{\mathcal{A}}[\Pr(\bar{S}_j=1\mid \bar{z}'_j,c(\bar{z}'_j), \bar{\gamma_j},\nu_j,\mathcal{A})]} - \nonumber \\  
& \quad \Delta_{\bar{z}'_{j}} \times \left(1-\frac{E_{\mathcal{A}}[\Pr(\bar{S}_j=1\mid \bar{z}_j,c(\bar{z}_j), \bar{\gamma_j},\nu_j,\mathcal{A})]}{E_{\mathcal{A}}[\Pr(\bar{S}_j=1\mid \bar{z}'_j,c(\bar{z}'_j), \bar{\gamma_j},\nu_j,\mathcal{A})]} \right),
\end{align}
where $\mathcal{A}$ denotes the set of temporally preceding variables $(\bar{y}_{j-1}, \bar{w}_{j}, \bar{r}_{j}, x_{0})$.
\item[Result 2:] $\tau$ in [\ref{Tau.eq}] can further be identified using Assumption 5 by weighting the contrasts in [\ref{sace.eq}] with  
\begin{align}\label{weights}
\frac{\Pr[\bar{S}_j(\bar{z}_j) = \bar{S}_j(\bar{z}'_j)= 1]}{\sum_{k=1}^J \Pr[\bar{S}_k(\bar{z}_k) = \bar{S}_k(\bar{z}_k^\prime)= 1]}=\frac{E_{\mathcal{A}}[\Pr(\bar{S}_j=1\mid \bar{z}_j,c(\bar{z}_j), \bar{\gamma_j},\nu_j,\mathcal{A})]}{\sum_{k=1}^J E_{\mathcal{A}}[\Pr(\bar{S}_k=1\mid \bar{z}_k,c(\bar{z}_k), \bar{\gamma_k},\nu_k,\mathcal{A})]}. 
\end{align}
\end{description}
The proofs of the results can be found in the Web Appendix section A.3. The causal effect is identifiable based on the observed data and Assumptions 1-7, conditional on the fixed values for the sensitivity parameters $c(\bar{z}_j)$, $c(\bar{z}'_j)$, $\Delta_{\bar{z}'_{j}}$, $\nu_j$ and, $\gamma_j$. For  a Bayesian analysis, the sensitivity parameters can be given informative priors. In Section 5.3 and Table \ref{algorithm} we describe the estimation algorithm where the sensitivity parameters are given informative, non-degenerate, priors.

\section{Modeling of the observed data distributions and computation of the causal effect} 
The joint distribution of the observed data 
is specified as a marginal model for the baseline confounders and a set of sequential conditional models for the time-varying variables, given the history of the joint process (the outcome, exposure, confounders, and missingness). Details of the joint distribution are given in Web Appendix section A.1. The baseline confounders $x_{i0}$ are all observed before an individual enters the study. For each visit $j$ we postulate the time-varying variables in the following order: $(s_{ij}, r_{ij},  w_{ij}, z_{ij}, y_{ij})$, even though the exposure, the time-varying confounder, and survival all occurred between $(j-1)$st and $j$th test wave. Of course, $y_{ij}$, $w_{ij}$ and $z_{ij}$, are only observed if $\bar{r}_{ij}=1$ and $\bar{s}_{ij}=1$. It is further allowed that $w_{ij}$ and $z_{ij}$ may have occurred before $s_{ij}$.

\subsection{Bayesian semi-parametric modeling}
We propose a Bayesian semi-parametric modeling approach based on Bayesian Additive Regression Trees (BART, \cite{chipman2010bart}) for the observed data distribution. 

For the time varying components, we specify BART models for the responses as a function of prior histories for all individuals alive and not dropped out at a given test wave. The model consists of two parts: a sum-of-trees model and a regularization prior on the parameters of that model. The model for the continuous response $Y_j$ is conditioned on the history of the joint process $(\bar{y}_{j-1}, \bar{z}_{j}, \bar{w}_{j}, x_0)$ for the subset that satisfies $\bar{r}_j=1$ and $\bar{s}_j=1$, and can be expressed as
$Y_j = \sum_{k=1}^{K_{Y_j}} g_{Y_j}\left((\bar{y}_{j-1}, \bar{z}_{j}, \bar{w}_{j},x_0);T_{Y_j}^k,M_{Y_j}^k\right) + \varepsilon_j.$
The model consists of $K_{Y_j}$ distinct binary regression trees denoted by $T_{Y_j}^k$. Each tree constitute a set of interior node decision rules leading down to $b_{Y_j}^k$ terminal nodes, and for a given $T_{Y_j}^k$, $M_{Y_j}^k =(\rho_{Y_j}^{k,1},\ldots,\rho_{Y_j}^{k,b^k})$ is the associated terminal node parameters. The conditional distribution of the continuous outcome is specified as normal, 
$Y_j \sim N\left(\mu_{Y_j}(\bar{y}_{j-1}, \bar{z}_{j}, \bar{w}_{j},x_0), \sigma_j^2\right),$
where the mean function, $\mu_{Y_j}(\bar{y}_{j-1}, \bar{z}_{j}, \bar{w}_{j},x_0)$, is given by the sum-of-trees.

The BART models for our binary responses $Z_j, W_j, R_j$, and $S_j$ are specified as probit models. For example the model for the exposure 
can be expressed as:
$\pi_{Z_j}(\bar{y}_{j-1}, \bar{z}_{j-1}, \bar{w}_{j}, x_0)$ \newline $= \Phi \left(
\sum_{k=1}^{K_{Z_j}} g_{Z_j}\left((\bar{y}_{j-1}, \bar{z}_{j-1}, \bar{w}_{j},x_0);T_{Z_j}^k,M_{Z_j}^k\right) \right),$ where $\Phi$ denotes the cumulative density function of the standard normal distribution 
and $\pi_{Z_j}(\bar{y}_{j-1}, \bar{z}_{j-1}, \bar{w}_{j-1}, x_0)$ is the probability of being exposed at wave $j$ given $(\bar{y}_{j-1}, \bar{z}_{j-1}, \bar{w}_{j}, x_0)$ for the subset that satisfies $\bar{r}_j=1$ and $\bar{s}_j=1$. The BART model for $S_j$ is fitted for the subset that satisfies $\bar{r}_{j-1}=1$ and $\bar{s}_{j-1}=1$, and for $R_j$ the subset that satisfies $\bar{r}_{j-1}=1$ and $\bar{s}_{j}=1$. The predicted probabilities of $r_j=1$ and $s_j=1$ are:
$\pi_{R_j}(\bar{y}_{j-1}, \bar{z}_{j-1}, \bar{w}_{j-1}, x_0)$
and 
$\pi_{S_j}(\bar{y}_{j-1}, \bar{z}_{j-1}, \bar{w}_{j-1}, x_0).$
Note that, $s_0=1$ and $r_0=1$ for all individuals, $\pi_{R_j}=0$  if $r_{j-1}=0$, and $\pi_{S_j}=0$ if $s_{j-1}=0$.

The baseline confounders are all categorical (age cohort, sex, and education level). We create a saturated multinomial random variable, $x_0\sim Multi(N, \pi_{x_0}^1, \pi_{x_0}^2, \ldots, \pi_{x_0}^L),$ based on these categorical variables.
$L$ is the number of categories and each category corresponds to a unique combination of the categorical variables. $\pi_{x_0}=(\pi_{x_0}^1, \pi_{x_0}^2, \ldots, \pi_{x_0}^L)$ is given a Dirichlet prior with parameters equal to one.
 
\subsection{Posterior}
Draws from the posterior distribution of the sum-of-trees models are generated using Markov chain Monte Carlo (MCMC). The parameters of the conditional distributions for $Y_j, Z_j, W_j, R_j$, and $S_j$ are assumed independent and thus their posteriors can be sampled simultaneously. BART is implemented in the R package \textit{bartMachine} (\cite{kapelner2013bartmachine}) for continuous and binary responses. We use default priors on all of the parameters of the sum-of-trees model, that is, on the tree structure, the terminal node parameters, and the error variance. For details see \citeauthor{kapelner2013bartmachine} (\citeyear{kapelner2013bartmachine}).

\subsection{Computation of the SACE}
The algorithm for generating samples from the posterior distribution of $\tau$ in [\ref{Tau.eq}] using the G-computation formula is given in Table \ref{algorithm}. Details can be found in the Web Appendix section A.4. The algorithm provides the details of generating posterior samples of the causal quantities in Results 1 and 2 (from Section 4) using the posterior distribution of the observed data model parameters (Section 5.1) and the identifying restrictions with sensitivity parameters (Sections 3.1 and 3.2).  Recall the expressions in Results 1 and 2 are a function of the observed data distribution and the sensitivity parameters.

For implementation of the algorithm in practice, a number of the initial posterior samples are discarded as burn-in. Parallel computation can be implemented to speed up computations. For example, instead of running one long chain in Step $1$, it is possible to run multiple shorter chains in parallel, although each chain still needs to converge. Also, Step $2$ may be divided into $k$ blocks of size $N^*/k$, and in Steps $3-4$ the parameters of interest are computed by combining the pseudo data from the $k$ blocks. We give further details on computation with Betula data in Section 7.3.

\section{Simulation study}
We performed a simulation study to evaluate the performance of the BSP G-computation algorithm. 
For simplicity of comparison to other appropriate methods we estimate $\mathrm{E}[Y_j(\bar{z}_j)-Y_j(\bar{z}'_j)\mid \bar{s}_j=1]$ and set $\Delta_{\bar{z}'_{j}}=0$, $\gamma_j=0$, and $c(z_j)=c(z'_j)=0$, i.e. a setting with MAR missingness and and no deaths. Details are found in the Web Appendix section A.5. 
 
We consider two settings for our BSP approach. First, where we specify a normal distribution for the outcome as described in the algorithm (BSP-GC1), and second (BSP-GC2), when specifying a t-distribution with 3 degrees of freedom ($t_3$). We compare our approach with three other methods used for causal effect estimation of longitudinal data with time-varying confounding. The three other methods implemented are: (i) A parametric version of the proposed procedure (BP-GC). Here we specified Bayesian linear and logistic additive regression models instead of the BART models described in Section 5.1. (ii) Inverse probability of treatment weights (IPTW; \cite{cole2008constructing}). Here, the mean $E[Y_j \mid \bar{s}_j=1, \bar{z}_j]$ is estimated by averaging the memory outcome for the subset with $\bar{Z}_j=\bar{z}_j$ in a pseudo-population constructed by weighting each individual using both unstabilized weights (IPTW-W) and stabilized weights (IPTW-SW), to adjust for confounding and for attrition among survivors. The IPTW-W and IPTW-SW were implemented using the \textit{ipw} and \textit{survey} packages in R.
(iii) Targeted minimum loss-based estimation approach for longitudinal data structures (TMLE; \cite{van2012targeted}). We implemented the TMLE using the \textit{ltmle} package using default settings (\cite{lendle2017ltmle}). Confidence intervals were calculated using nonparametric bootstrap. We used 5000 bootstrap samples. The bootstrap confidence intervals were calculated using the 2.5th and 97.5th percentiles of the resulting estimates. 

Data were generated based on a simplified version of the Betula data. We simulated 1000 datasets of size n = 1000. We considered $J_i = 2$ follow-up test waves, a continuous baseline covariate, $X_{i0}$, generated as $X_0 \sim Unif\left( 0, 1 \right)$. The outcome, $Y_{ij}$, was considered a continuous time-varying variable. The binary variable $Z_{ij}$ indicated if the subject was widowed or not, and  $W_{ij}$ indicated if the spouse been severely sick. Widowhood was an absorbing state, such that, if $Z_{ij}=1$ then $Z_{ik}=1$ for $k \geq j$. Note, that $Z_{i0}=0$ for all subjects. As in the Betula data, all time-varying variables had a highly nonlinear relationship with the baseline covariate and the time-varying confounder interacted with the baseline covariate in the exposure model. 
Data for the simulation study was generated as $ X_0 \sim Unif\left( 0, 1 \right)$, 
$W_j \sim $ $Ber\left(expit(-2 + 0.5 X_0 - 2 X_0^2 + 5 X_0^3 + 0.25 W_{j-1}) \right)$, 
\\$Z_j  \sim$ $Ber\left(expit(-5 + X_0 - 4 X_0^2 + 6 X_0^3 + 0.6 W_{j} + 0.3 W_{j-1} + 0.5 X_0 W_{j} - X_0^2 W_{j} + 2 X_0^3 W_{j})\right)$, and 
$Y_j  = 0.5 - 0.05 Z_j  - 0.05 W_j  - 0.25 Y_{j-1} - 0.1 X_0 + 0.25 X_0^2 - 0.25 X_0^3  +  \epsilon_j$, where $\epsilon_j \sim N\left(0, 0.1^2\right)$.
R code for the data generation is provided in the Web Appendix section A.5.

Table \ref{simulation.study} shows the bias, empirical standard deviation (ESD), mean squared error (MSE), and coverage of $95\%$ confidence intervals from the simulation study for BSP-GC1, BSP-GC2, BP-GC, IPTW-W, IPTW-SW, and TMLE. The causal effect estimates for BSP-GC1, BSP-GC2 and TMLE are nearly unbiased. BSP-GC1 and BSP-GC2 are however more efficient (smaller MSE and ESD) and have higher coverage (larger than $95\%$) than TMLE (lower than $95\%$). The simulation results for BSP-GC1 and BSP-GC2 are very similar. As expected, the three other methods; BP-GC, IPTW-W, and IPTW-SW are all biased. Additionally, of all methods, BP-GC has the highest bias and MSE, IPTW-W is the least efficient, and IPTW-SW has the lowest coverage.

To see how the proposed approach performs when the error distribution is misspecified data were instead generated from a $t_3$-distribution for the error of the outcome $Y_j$ in [\ref{data_generation_simulation}]. Bias, ESD, MSE, and coverage from the simulation are found in Table \ref{simulation.study.t3}. The results are similar in terms of bias and coverage compared to the previous simulation with correctly specified error. However, ESD and MSE are higher and are now comparable to TMLE.

To see how the proposed approach performs when there is lack of overlap, data were generated as $Ber\left(expit(a_{Z_j})\right) I_{X_0>0.5}$ for the exposure, mimicking the Betula study where only individuals at older ages were exposed (widowed). The results from the simulation are found in Table \ref{simulation.study.t3}. The results are similar to the first simulation with correctly specified error and non-linear effects for BSP-GC1 and TMLE. But for the other three approaches bias was higher and CP was lower.

\section{Analysis of the Betula data}

\subsection{The Betula data}
The goal is to estimate the causal effect of becoming a widow on memory among those who would survive irrespective of being widowed or not. As such, we limit our data set to those individuals who were married at enrollment. Of approximately 2000 participants $N=1059$ were married at study enrolment, and data were recorded at 4 fixed test waves ($j=0,\ldots, 3$) with 5 years interval. The memory outcome was assessed at each wave using a composite of three episodic memory tasks. The score can range between 0 and 76, with a higher score indicating better memory (for details see \cite{josefsson12}). We consider two contrasting exposure regimes, subjects who became a widow between the $j-1$th and $j$th wave, $\bar{z}_j=\{z_0=0,\ldots,z_{j-1}=0,z_j=1\}$, and subjects married through test wave $j$, $\bar{z}'_j=\{z_0=0,\ldots,z_j=0\}$, for $j=1,2,3$. Baseline demographic characteristics included age-cohorts: $45, 50, \ldots , 80$ years of age at enrollment, gender, and education, categorized into \textit{low}: 6-7 years of education (29\%), \textit{intermediate}: 8-9 years (31\%), or \textit{high}: \textgreater{9} years (40\%). We also measured a time-varying confounder; an indicator if the spouse has been sick within the last 5 years. We note that baseline confounders are always recorded.

\subsection{Sensitivity parameters}
Our approach allows uncertainty about untestable assumptions by specifying priors for the sensitivity parameters described in Section 3.2. We restrict the parameters to a plausible range of values, reflecting the authors' beliefs about the unknown quantities. 

In Section 3.2, the sensitivity parameter $c(\bar{z}_j)$ reflects the average difference in potential outcomes due to unmeasured confounding (violation of Assumption 3). For the Betula data, when studying the effect of widowhood on cognition, one concern may be that the association is confounded by a healthy lifestyle, such as a healthy diet and/or exercise, something that is often shared within couples. Couples with a healthy lifestyle live longer and may have better cognitive performance than couples with a less healthy lifestyle. This information is not available from the database. Hence, it is a potential unmeasured confounder. Here, we assume $c(z_j)<0$ and $c(z'_j)>0$, reflecting that exposed (widowed) individuals are less healthy compared to unexposed (married) individuals. We further assume the effect is equal for exposed and unexposed. That is, we assume $c(z_j)=-\xi_j$ and $c(z'_j)=\xi_j$. Here, we specify a uniform prior on the sensitivity parameters, $\xi_j\sim \mathrm{Unif}(0, U_{\xi_j}),$ with upper bound $U_{\xi_j}= \frac{1}{2}\times SD(Y_j \mid \bar{y}_{j-1}, \bar{z}_{j}, \bar{w}_{j}, \bar{r}_j=1, \bar{s}_j=1, x_0).$ That is, we expect the sensitivity parameter not to be bigger than one-half standard deviation of the outcome conditional on the history of the joint process. This approximately corresponds to an effect size similar to that found in previous literature on the effect of Mediterranean diet on memory (\cite{radd2018effect}).

Departures from a MAR mechanism (Assumption 4) for the missingness among survivors can be investigated by varying $\gamma_j$ in Section 3.2. Our prior belief is that $\gamma_j<0$, reflecting a negative shift in memory performance occur immediately after the first unobserved test wave. Here, the prior is specified as $\gamma_j\sim \mathrm{Unif}(-L_{\gamma_j}, 0),$
where we assume the lower bound is one observed conditional standard deviation,
$L_{\gamma_j} = 1 \times SD(Y_j \mid \bar{y}_{j-1}, \bar{z}_{j}, \bar{w}_{j}, \bar{r}_j=1, \bar{s}_j=1, x_0).$ 
The effect is similar to what has been found in previous work examining differences in cognition between completers and those who withdraw, at the last cognitive testing visit before dropping out (\cite{rabbitt2008death}). 

Sensitivity to Assumption 6, uses $\Delta_{\bar{z}'_{j}}$, which reflects the difference in outcomes when comparing the "always survivor" strata to the strata where individuals were to live under the contrasting regime $\bar{z}'_{j}$ but not under exposure regime $\bar{z}_{j}$. We again specify a uniform prior $\Delta_{\bar{z}'_{j}} \sim \mathrm{Unif}(0, U_{\Delta_{\bar{z}'_{j}}}),$
with upper bound $U_{\Delta_{\bar{z}'_{j}}}=1\times SD(Y_j \mid \bar{s}_j=1).$ 

Finally, sensitivity to Assumption 7 uses the sensitivity parameter $\nu_j$, which represents the difference in the probability of being exposed at wave $j$ for non-survivors and survivors conditioning on the history of the joint process. As shown in Section 3.2, $\nu_j$ is restricted to $[0, U_{\nu_j}]$. We assume the prior for $\nu_j$ is uniform over this range, $\nu_j \sim \mathrm{Unif}(0, U_{\nu_j}).$ The upper bound reflects that, between the $j-1$th and $j$th wave, all subjects were exposed before death.

\subsection{Results and comparison with other methods}
We estimated $\tau$ using the proposed BSP method and embedded sensitivity parameters. For each chain the first 1000 iterations were discarded as burn-in, and 2240 posterior samples of $\tau$ were obtained. We sampled pseudo data of size $N^*=25000$ at each iteration. Convergence of the posterior samples was monitored using trace plots of the samples. To reduce computation time we used 448 parallel chains. Total computation time was 1 hour and 18 minutes.

For longitudinal exposure regimes limited overlap is not uncommon. To avoid extrapolation of the outcome model outside the range of estimated propensities we restrict the overlap region for the longitudinal exposure regimes. Specifically, we restrict data to the set of individuals that have an estimated propensity score that lies within the range of the observed propensities for the two contrasting regimes (similar to the procedure used in \cite{Zhou2019positivity}).

We consider two settings for our BSP approach. First, we specify a normal distribution for the residual of the outcome as described in the algorithm (BSP-GC1); second (BSP-GC2), we replace the normal distribution with a t-distribution with 3 degrees of freedom ($t_3$). For BSP-GC1, the posterior sampling results revealed a mean episodic memory score of 38.2 (95\% CI; $35.4, 40.8$) for exposed and 38.1 (95\% CI; $35.6, 40.1$) for unexposed individuals, and an estimate of $\tau$ of 0.18 (95\% CI; -$1.43,1.86$), suggesting that there is no effect of becoming a widow on memory among those who would survive irrespective of exposure. For BSP-GC2, the posterior sampling results revealed a mean episodic memory score of 38.2 (95\% CI; $35.5, 40.9$) for exposed and 38.0 (95\% CI; $35.6, 40.1$) for unexposed individuals, and an estimate of $\tau$ of 0.21 (95\% CI; -$1.42,1.82$). The conclusions are insensitive to the two choices of outcome residual distribution here.

As a sensitivity analysis we compare how the point estimates and uncertainty varied when setting one sensitivity parameter at a time to zero, while the remaining sensitivity parameters are given the priors described in Section 7.2. Setting $\gamma_j$ to zero resulted in a estimate of $\tau$ of $0.18$ (95\% CI; $-1.42, 1.91$); for $\nu_j=0$, $0.20$ (95\% CI; $-1.36, 1.83$); for $\delta=0$, $0.21$ (95\% CI; $-1.38, 1.94$); and for $\xi_j=0$, $-0.83$ (95\% CI; $-2.43, 0.75$). The largest differences was found for the analysis setting $\xi_j$ to zero (i.e. no unmeasured confounding); however the CI still cover zero and we expect this assumption to not hold. Fixing the other sensitivity parameters at zero had minimal impact.  

We also compare our approach, BSP-GC1 with BP-GC, IPTW-W, IPTW-SW, and TMLE (described in the simulation study). For simplicity of comparison we estimate the causal contrasts described in the simulation study. Further to avoid limited overlap, we restrict our data to those age-cohorts where we observe both married and widowed participants over the study period, instead of restricting to the region as for the main analyses. For IPTW-W, IPTW-SW, and TMLE, confidence intervals were calculated using nonparametric bootstrap. We used 5000 bootstrap samples. The bootstrap confidence intervals were calculated using the 2.5th and 97.5th percentiles of the resulting estimates.   

The results from all the methods are given in Table \ref{comparison.otherm}. First, all of the methods display a negative widowhood effect on memory, although all confidence/credible intervals (CI) cover zero. There is a large discrepancy between our semi-parametric approach, BSP-GC1, and the parametric counterpart, BP-GC. In the latter, the effect was attenuated and the CI was narrower. A likely explanation of the discrepancy in effect estimates is that BP-GC is more susceptible to bias caused by model misspecification. BP-GC and IPTW-SW yielded most similar results, although the weighting approach had much wider CI. Further, the effect estimate appeared most negative using IPTW-W and the CI was much wider than for any of the other methods. Weighting methods are known to be unstable and to have problems with large variance estimates in finite samples if the values of the weights are extreme. In our analysis the range of the weights was 0.06-14.3 for IPTW-W, compared to 0.06-5.4 for IPTW-SW. The large weights using IPTW-W may explain the deviating result using this method. Our BSP-GC1 approach yielded an estimate of $\tau$ most similar to TMLE, although TMLE had slightly wider CI. This is consistent with the results of the simulation study.
\section{Concluding remarks}
This paper has proposed a Bayesian semi-parametric (BSP) framework for estimating the SACE with longitudinal cohort data. Our approach allows for Bayesian inference under MNAR missingness and truncation by death, as well as the ability to characterize uncertainty about unverifiable assumptions. The proposed approach has several advantages compared to existing approaches: (i) the flexible modeling of the observed data as compared to parametric methods, while maintaining computational ease, (ii) interval estimates for full posterior inference, (iii) easy to introduce sensitivity parameters. 

The simulation study, although simplified, mirrored the Betula data. All time-varying variables had a highly nonlinear relationship with the baseline covariate and interaction effects were included. The models for BP-GC, IPTW-W, IPTW-SW, and TMLE were specified using additive effects, and thus, were misspecified. The results showed that BSP-GC1, BSP-GC2 and TMLE were nearly unbiased. BSP-GC1 and BSP-GC2 were however more efficient and had better coverage than TMLE (though a bit conservative). The results are in line with previous research (\cite{roy2018bayesian}), suggesting that TMLE is less efficient than Bayesian semi-parametric and non-parametric modeling. This, however, must be explored more thoroughly in future work.

The three other methods; BP-GC, IPTW-W, and IPTW-SW, were all biased. This is expected since these methods make stronger distributional assumptions and thus are more sensitive to model misspecification. Similar to TMLE our approach does not rely on strong modeling assumptions, but unlike TMLE, it is quite easy to modify assumptions and incorporate sensitivity parameters. Recall we could not easily make direct comparisons of the proposed approach with the other approaches under our assumptions that include sensitivity parameters. We attempted to implement Super learner, implemented in the R package \textit{SuperLearner}, but observed highly variable results for the Betula data (causal effect estimates varied between -0.34 and -1.33). This may be a result of the cross-validation step and the fact that the exposure is a rather rare event. Using our BSP approach these problems are avoided by increasing the size of the pseudo data and running longer chains. Although, computation time can be demanding for large pseudo sample sizes, the algorithm can be fully parallelized as discussed in Section 5.3 and Section 7.3, which would vastly reduced the total computation time. 

For the Betula data we did not find an effect of widowhood on memory. The results were not sensitive to different specification of the errors as normal- or t-distributed, and changing the sensitivity parameters one at a time did not change the results significantly either. The difference in findings from previous studies may partly be explained by different estimands being used; ours is the only analysis using a SACE. Additionally, in this study we considered the immediate effect of widowhood (within 5 years) rather than a long term effect; it may take longer for degeneration to become apparent. 

Our approach can be generalized in various ways. For example, it is possible to allow for multiple time-varying confounders and/or continuous baseline confounders using a sequential approach as proposed by (\cite{xu2016sequential}). This would involve first ordering the confounders into sequential conditionals and then applying BART to model each of these univariate conditionals. Additionally, although widowhood status is thought of as a monotone exposure pattern and an absorbing state in this study, this is not essential for the proposed approach and other (non-monotone) exposure regimes, such as the effect of widowhood duration on memory at the last visit, might be of interest and are possible to study with a few modifications (for example, the positivity assumption).

Violations of the consistency assumption can be problematic when using observational data (\cite{cole2009consistency}). For example, the effect of widowhood can affect memory via different pathways, e.g. for some subjects via stress or depression and others via reduced physical health due to poorer lifestyle choices (\cite{gerritsen2017influence}). This is a limitation with the current study where the exposure is defined homogeneously, and should be explored more thoroughly in future work.

Several of our assumptions can be (further) relaxed. For example, Assumption 5 can be weakened to a stochastic Monotonicity, by following the procedure described in \citeauthor{lee2010causal} (\citeyear{lee2010causal}). Also, in this study we have considered unmeasured outcome confounding; this assumption can easily be extended to allow unmeasured mortality confounding. Assumption 6 can be weakened by conditioning on the history of the joint process. However, a drawback with relaxing these assumptions is increasing the number of sensitivity parameters. 

One limitation with BART is the restrictive, and sometimes unrealistic, assumption of IID normal errors, (although they can easily be replaced with heavier tail errors as in here). A fully non-parametric modeling approach could be obtained by extending BART to model the error distribution using the Dirichlet process mixtures (\cite{george2018fully}). An additional limitation of the proposed approach is that we used existing R-functions for BART that are not most efficient for our setting. We will explore these limitations in future work, as well as, other choices for priors of the sensitivity parameters. 

\section*{Supplementary materials}
Web Appendices referenced in Sections 3, 4, 5, 6, and 7, as well as R code are available as Supplementary materials. 

\section*{Acknowledgments}
The authors would like to thank Dr Anna Sundstr{\"o}m for helpful discussions on the interpretation of the results.
This work is partially funded by The Swedish Foundation for Humanities and Social Sciences P17-0196:1 and   Paths to Healthy and Active Ageing, funded by the Swedish Research Council for Health, Working Life and Welfare, (Dnr 2013 – 2056) to MJ.  This work is partially funded by US NIH grants CA183854 and GM112327 to MJD. This publication is based on data collected in the Betula prospective cohort study, Umeå University, Sweden. The Betula Project is supported by Knut and Alice Wallenberg foundation (KAW) and the Swedish Research Council (K2010-61X-21446-01). The simulations were enabled by resources provided by the Swedish National Infrastructure for Computing (SNIC) at Umea University partially funded by the Swedish Research Council through grant agreement no. 2016-07213.

\printbibliography

\newpage
\section{Tables and Figure}
\begin{table}[hp]
\caption{The table shows possible missing data, $\bar{R}$, and mortality patterns, $\bar{S}$. The outcome vector $\mathbf{Y}=\left\lbrace Y_0, Y_1, Y_2, Y_3 \right\rbrace$ is fully observed if $\bar{S}=\bar{R}=1$, otherwise it is constrained by the mortality outcome and/or missing data patterns. $Y_j=\mathrm{O}$ if the outcome is observed, $Y_j=\mathrm{M}$ if missing, and $Y_j=\mathrm{nd}$ when truncated by death. The NFD-S restriction leaves the distribution for $Y_j=\mathrm{M}^*$ unidentified.} 
\label{dropoutmortality.pat}
\begin{adjustbox}{width=1\textwidth}
    \begin{tabular}{ c  c  c  c  c }
    \hline
    & \multicolumn{4}{ c }{$\bar{R}_J$} \\ 
    \hline
 $\bar{S}_J$    & $\{1,0,0,0 \}$ & $\{1,1,0,0 \}$ & $\{1,1,1,0 \}$ & $\{1,1,1,1 \}$ \\ 
\hline
$\{1,0,0,0 \}$ & $\left\lbrace  \mathrm{O},\mathrm{nd},\mathrm{nd},\mathrm{nd} \right\rbrace$ & - & - & - \\ 
$\{1,1,0,0 \}$ & $\left\lbrace  \mathrm{O},\mathrm{M}^*,\mathrm{nd},\mathrm{nd}\right\rbrace$ & $\left\lbrace  \mathrm{O},\mathrm{O},\mathrm{nd},\mathrm{nd}\right\rbrace$ & - & - \\ 
$\{1,1,1,0 \}$ & $\left\lbrace  \mathrm{O},\mathrm{M}^*,\mathrm{M},\mathrm{nd}\right\rbrace$ & $\left\lbrace  \mathrm{O},\mathrm{O},\mathrm{M}^*,\mathrm{nd}\right\rbrace$ & $\left\lbrace  \mathrm{O},\mathrm{O},\mathrm{O},\mathrm{nd}\right\rbrace$ & - \\ 
$\{1,1,1,1 \}$ & $\left\lbrace  \mathrm{O},\mathrm{M}^*,\mathrm{M},\mathrm{M}\right\rbrace$ & $\left\lbrace  \mathrm{O},\mathrm{O},\mathrm{M}^*,\mathrm{M}\right\rbrace$ & $\left\lbrace  \mathrm{O},\mathrm{O},\mathrm{O},\mathrm{M}^*\right\rbrace$ & $\left\lbrace  \mathrm{O},\mathrm{O},\mathrm{O},\mathrm{O}\right\rbrace$ \\       
\hline
\end{tabular}
\end{adjustbox}
\end{table}

\begin{table}[hp]
\caption{Algorithm for estimation of $\tau$ in [\ref{Tau.eq}] using the G-computation formula. Details of the algorithm can be found in the Web Appendix section A.4.} 
\label{algorithm}
\begin{tabular}{ l l}
\hline
1. & Sample the observed data posteriors as described in Section 5. \\
2. & For each posterior sample of the parameters sample pseudo data $(\bar{y}^*_{j-1},\bar{w}^*_{j},\bar{r}_{j}^*,\bar{s}_{j}^*,x_0^*)$ \\
& and sensitivity parameters $\gamma_j$, $\nu_j$, $c(z_j)$, and $c(z'_j)$ of size $N^*$. Additionally, sample \\
& one set of $\Delta_{\bar{z}'_{j}}$.\\
3. & Implement G-computation for $\bar{z}_j$, and similarly for $\bar{z}'_j$, using the pseudo data and \\
& sensitivity parameters from Step 2 by computing $E[Y_j \mid \bar{y}_{j-1}, \bar{z}_j, \bar{w}_{j},\bar{r}_{j}, \bar{s}_j=1, x_0]$\\
&  and $\prod^j_{k=0} \Pr[S_{k}=1 \mid \bar{z}_{k}, \bar{w}_k,\bar{r}_k,\bar{y}_{k-1},\bar{S}_{k-1}=1, x_0]$. Furthermore, implement Monte \\
& Carlo integration using the pseudo data to compute $\Pr[\bar{S}_{j}=1 \mid \bar{z}_j]$ and  \\
& $E[Y_j, \bar{S}_j=1 \mid \bar{z}_j]$. \\
4. & Use the quantities in Step 3 to compute one posterior sample of $\tau$ as defined in [\ref{sace.eq}]-[\ref{weights}]. \\ 
5 & Repeat step 2 - 4 for each of the posterior sample of the parameters.  \\ 
\hline
\end{tabular}
\end{table}

\begin{table}[hp]
\caption{Simulation results for causal effect estimation with n=1000 with the true causal effect $\tau=-0.05$, using two settings for our proposed approach: error specified as normal (BSP-GC1) and error specified using a t-distribution (BSP-GC2), a parametric version of the proposed procedure (BP-GC), inverse probability of treatment weights using unstabilized weights (IPTW-W) and stabilized weights (IPTW-SW), and Targeted minimum loss-based estimation approach for longitudinal data structures (TMLE). Mean squared error (MSE) are multiplied by 100 for ease of presentation. ESD denotes empirical standard deviation and CP denotes coverage probability of 95\% credible intervals.
} 
\label{simulation.study}
\begin{center}
    \begin{tabular}{l c c c c}
    \hline
    & Bias & ESD & MSE & CP\\ 
\hline
BSP-GC1 & -0.002 & 0.013 & 0.02 & 98.7\\
BSP-GC2 & -0.003 & 0.013 & 0.02 & 98.4\\
BP-GC & -0.065 & 0.021 & 0.47 & 63.6 \\
IPTW-W & 0.024 & 0.040 & 0.22 & 64.3 \\ 
IPTW-SW & -0.031 & 0.013 & 0.12 & 20.4 \\ 
TMLE & -0.002 & 0.021 & 0.04 & 92.8 \\ 
\hline
\end{tabular}
\end{center}
\end{table}

\begin{table}[hp]
\caption{Simulation results for causal effect estimation with n=1000 with the true causal effect $\tau=-0.05$. Two scenarios, a) where the error distribution for the outcome is misspecified using a t-distribution with 3 degrees of freedom and b) a setting with limited overlap. Comparing our proposed approach (BSP-GC1), a parametric version of the proposed procedure (BP-GC), inverse probability of treatment weights using unstabilized weights (IPTW-W) and stabilized weights (IPTW-SW), and Targeted minimum loss-based estimation approach for longitudinal data structures (TMLE). Mean squared error (MSE) are multiplied by 100 for ease of presentation. ESD denotes empirical standard deviation and CP denotes coverage probability of 95\% credible intervals.
} 
\label{simulation.study.t3}
\begin{center}
    \begin{tabular}{l c c c c c c c c }
    \hline
    & \multicolumn{4}{l}{a) $t_3$} & \multicolumn{4}{l}{b) non-overlap} \\
    & Bias & ESD & MSE & CP & Bias & ESD & MSE & CP \\ 
\hline
BSP-GC1 & -0.002 & 0.037 & 0.13 & 97.2 & -0.003 & 0.014 & 0.02 & 98.9 \\
BP-GC & -0.074 & 0.038 & 0.69 & 69.5 & -0.078 & 0.024 & 0.67 & 43.0 \\
IPTW-W & 0.028 & 0.068 & 0.55 & 59.7 & -0.026 & 0.033 & 0.18 & 56.2 \\ 
IPTW-SW & -0.031 & 0.022 & 0.15 & 43.5 & -0.042 & 0.015 & 0.20 & 10.6 \\ 
TMLE & -0.001 & 0.036 & 0.13 & 92.3 & 0.002 & 0.023 & 0.05 & 93.9 \\ 
\hline
\end{tabular}
\end{center}
\end{table}

\begin{table}[hp]
\caption{Comparison of methods used for causal effect estimation of the Betula data, setting $\Delta_{\bar{z}'_{j}}=0$, $\gamma_j=0$, and $c(z_j)=c(z'_j)=0$, using our proposed approach (BSP-GC), a parametric version of the proposed procedure (BP-GC), inverse probability of treatment weights using unstabilized weights (IPTW-W) and stabilized weights (IPTW-SW), and Targeted minimum loss-based estimation approach for longitudinal data structures (TMLE).
} 
\label{comparison.otherm}
\begin{center}
    \begin{tabular}{ l  c }
    \hline
     & Estimate [95\% CI] \\ 
\hline
BSP-GC & -0.98 [-2.78, 0.73] \\
BP-GC & -0.53 [-1.73, 0.68] \\
IPTW-W & -1.67 [-5.96, 1.51] \\ 
IPTW-SW & -0.44 [-3.06, 1.39] \\ 
TMLE & -0.96 [-3.11, 0.99] \\ 
\hline
\end{tabular}
\end{center}
\end{table}

\end{document}


\maketitle
\section*{A.1: Details on the joint distribution of the observed data}
We specify a marginal model for the baseline confounders and a set of sequential conditional models for the time-varying variables, given the history of the joint process (the outcome, exposure, confounders, and missingness) as follows:
\begin{align*}
\prod_{k=J^r_i+1}^{J^s_i}  & p(s_{ij} \mid \bar{y}_{iJ^r_i}, \bar{z}_{iJ^r_i}, \bar{w}_{iJ^r_i}, \bar{r}_{iJ^r_i}=1, r_{iJ^r_i+1}=0,\ldots , r_{iJ^s_i}=0, \bar{s}_{ij-1}=1, x_{i0}) \nonumber \\ 
\prod_{j=0}^{J^r_i}  & \Bigl(p(y_{ij} \mid \bar{y}_{ij-1}, \bar{z}_{ij}, \bar{w}_{ij}, \bar{r}_{ij}=1, \bar{s}_{ij}=1, x_0) \times \nonumber \\ 
& p(z_{ij} \mid \bar{y}_{ij-1}, \bar{z}_{ij-1}, \bar{w}_{ij}, \bar{r}_{ij}=1, \bar{s}_{ij}=1, x_{i0}) \times \nonumber \\ 
& p(w_{ij} \mid \bar{y}_{ij-1}, \bar{z}_{ij-1}, \bar{w}_{ij-1}, \bar{r}_{ij}=1, \bar{s}_{ij}=1, x_{i0}) \times \\ 
& p(r_{ij} \mid  \bar{y}_{ij-1}, \bar{z}_{ij-1}, \bar{w}_{ij-1}, \bar{r}_{ij-1}=1, \bar{s}_{ij}=1, x_{i0}) \times \nonumber \\ 
& p(s_{ij} \mid \bar{y}_{ij-1}, \bar{z}_{ij-1}, \bar{w}_{ij-1}, \bar{r}_{ij-1}=1, \bar{s}_{ij-1}=1, x_{i0})\Bigr) \times \nonumber  \\ 
& p(x_{i0}).  \nonumber
\end{align*}

\section*{A.2: Derivation of bounds for $\mathbf{\nu_j}$ in Assumption 7.}
Here, we derive bounds for $\nu_j$. By using the law of total probability we have that
\begin{align*}
&\Pr[z_j \mid \bar{y}_{j-1},\bar{z}_{j-1},\bar{w}_j,r_j=0,\bar{s}_{j-1}= 1, x_0] \\
& = (\nu_j - \Pr[z_j \mid \bar{y}_{j-1},\bar{z}_{j-1},\bar{w}_j,\bar{r}_{j},\bar{s}_{j}= 1, x_0]) \\
& \quad \times \Pr[S_j=0 \mid \bar{y}_{j-1},\bar{z}_{j-1},\bar{w}_j,r_j=0,\bar{s}_{j-1}= 1, x_0] \\
& \quad + \Pr[z_j \mid \bar{y}_{j-1},\bar{z}_{j-1},\bar{w}_j,r_j=0,\bar{s}_{j}=1, x_0] \\
& \quad \times \Pr[S_{j}=1 \mid \bar{y}_{j-1},\bar{z}_{j-1},\bar{w}_j,r_j=0,\bar{s}_{j-1}= 1, x_0]\quad 
\end{align*}
and by using Bayes theorem we have that
\begin{align*}
&\Pr[z_j\mid  \bar{y}_{j-1},\bar{z}_{j-1},\bar{w}_j,r_j=0,\bar{s}_{j-1}= 1, x_0] \\
& = \Pr[z_j\mid  \bar{y}_{j-1},\bar{z}_{j-1},\bar{w}_j,r_j=0,\bar{s}_{j}= 1, x_0] \\ 
& \quad \times \frac{\Pr[S_{j}=1 \mid \bar{y}_{j-1},\bar{z}_{j-1},\bar{w}_j,r_j=0,\bar{s}_{j-1}= 1, x_0]}{\Pr[S_{j}=1 \mid \bar{y}_{j-1},\bar{z}_{j},\bar{w}_j,r_j=0,\bar{s}_{j-1}= 1, x_0]}.
\end{align*}
Plugging in the second equation for the right hand side and solving the first equation for $\nu_j$ leads to
\begin{align*}
\nu_j &= \Pr[z_j\mid \bar{y}_{j-1}, \bar{z}_{j-1},\bar{w}_j,r_j=0,\bar{s}_{j}= 1, x_0] \\ 
& \quad \times \frac{\Pr[S_{j}=1 \mid \bar{y}_{j-1},\bar{z}_{j-1},\bar{w}_j,r_j=0,\bar{s}_{j-1}= 1, x_0]}{\Pr[S_{j}=0 \mid \bar{y}_{j-1},\bar{z}_{j-1},\bar{w}_j,r_j=0,\bar{s}_{j-1}= 1, x_0]} \\
& \quad \times \frac{\Pr[S_{j}=0 \mid \bar{y}_{j-1},\bar{z}_{j},\bar{w}_j,r_j=0,\bar{s}_{j-1}= 1, x_0]}{\Pr[S_{j}=1 \mid \bar{y}_{j-1},\bar{z}_{j},\bar{w}_j,r_j=0,\bar{s}_{j-1}= 1, x_0]} \\
& \quad + \Pr[z_j \mid \bar{y}_{j-1},\bar{z}_{j-1},\bar{w}_j,\bar{r}_{j},\bar{s}_{j}= 1, x_0]).
\end{align*}
Further, by Assumptions 1 and 6 we have that 
\begin{align*}
& \Pr[S_{j}=1 \mid \bar{y}_{j-1},\bar{z}_{j},\bar{w}_j,r_j=0,\bar{s}_{j-1}= 1, x_0] \\
& \leq \Pr[S_{j}=1 \mid \bar{y}_{j-1},\bar{z}'_j,\bar{w}_j,r_j=0,\bar{s}_{j-1}= 1, x_0].
\end{align*} 
Thus, the lower bound for $\nu_j$ is obtained when $\Pr[S_{j}=1 \mid \bar{y}_{j-1},\bar{z}_{j},\bar{w}_j,r_j=0,\bar{s}_{j-1}= 1, x_0] = \Pr[\bar{S}_{j}=1 \mid \bar{y}_{j-1}, \bar{z}'_j,\bar{w}_j,r_j=0,\bar{s}_{j-1}= 1, x_0]$. Moreover, since the two regimes only differ in $z_j$, we note that,
\begin{align*}
&\Pr(\bar{S}_{j}=1\mid \bar{y}_{j-1},\bar{z}_{j-1},\bar{w}_j,r_j=0,\bar{s}_{j-1}= 1, x_0] \\
& = \Pr[\bar{S}_{j}=1 \mid \bar{y}_{j-1},\bar{z}_{j},\bar{w}_j,r_j=0,\bar{s}_{j-1}= 1, x_0] \\
& \quad \times \Pr[z_j\mid \bar{y}_{j-1},\bar{z}_{j-1},\bar{w}_j,r_j=0,\bar{s}_{j-1}= 1, x_0] \\
& \quad + \Pr[\bar{S}_{j}=1 \mid \bar{y}_{j-1}, z'_j,\bar{z}_{j-1},\bar{w}_j,r_j=0,\bar{s}_{j-1}= 1, x_0] \\
& \quad \times \Pr[z'_j\mid \bar{y}_{j-1},\bar{z}_{j-1},\bar{w}_j,r_j=0,\bar{s}_{j-1}= 1, x_0]\\
&= \Pr[\bar{S}_{j}=1 \mid \bar{y}_{j-1},\bar{z}_j,\bar{w}_j,r_j=0,\bar{s}_{j-1}= 1, x_0] 
\end{align*}
Hence, the lower bound for $\nu_j$ is
\begin{align*}
L_{\nu_j} = 0,
\end{align*}
and the upper bound for $\nu_j$ is
$$U_{\nu_j}=1-\Pr[z_j\mid  \bar{z}_{j-1},\bar{w}_j,\bar{R}_j=1,\bar{y}_{j-1},\bar{s}_{j}= 1, x_0].$$

\section*{A.3: Details on Equations [3] and [4] in Section 4.}
Here, we present results for identification of the causal estimand in [2] using Assumptions 1-8 introduced in Section 3. First, for identification of the contrasts $\mathrm{E}[Y_j(\bar{z}_j)-Y_j(\bar{z}'_j)\mid \bar{S}_j(\bar{z}_j) = \bar{S}_j(\bar{z}'_j)= 1]$ for $j=1,\ldots,J$ in [1], by using the law of total probability (ltp) and some algebra, we have that 
\begin{align*}
E&[Y_j(\bar{z}_j) \mid \bar{S}_j(\bar{z}_j)=1] \\
= & E[Y_j(\bar{z}_j) \mid \bar{S}(\bar{z}_j)=\bar{S}_j(\bar{z}'_j)=1] + \Pr[\bar{S}_j(\bar{z}'_j)\neq 1 \mid \bar{S}_j(\bar{z}_j)=1]\\
& \times (E[Y_j(\bar{z}_j) \mid \bar{S}_j(\bar{z}_j)=1, \bar{S}_j(\bar{z}'_j)\neq 1] - E[Y_j(\bar{z}_j) \mid \bar{S}(\bar{z}_j)=\bar{S}_j(\bar{z}'_j)=1]). 
\end{align*}
Using Assumption 7 and solving the above equation for $E[Y_j(\bar{z}_j) \mid \bar{S}_j(\bar{z}_j)=\bar{S}_j(\bar{z}'_j)=1]$ we obtain
\begin{align}\label{Solv}
&E[Y_j(\bar{z}_j) \mid \bar{S}(\bar{z}_j)=\bar{S}_j(\bar{z}'_j)=1]  \nonumber \\
&=E[Y_j(\bar{z}_j) \mid \bar{S}_j(\bar{z}_j)=1] + \Delta_{\bar{z}_j}  \Pr[\bar{S}_j(\bar{z}'_j)\neq 1 \mid \bar{S}_j(\bar{z}_j)=1]. \tag{A.1}
\end{align}
For exposure regime $\bar{z}_j$ (and similarly for $\bar{z}'_j$) and using Assumption 1, we have that
\begin{align}\label{expYS}
E[Y_j(\bar{z}_j) \mid \bar{S}_j(\bar{z}_j)=1]& = \frac{E[Y_j(\bar{z}_j), \bar{S}_j(\bar{z}_j)=1]}{\Pr[\bar{S}_j(\bar{z}_j)=1]} \nonumber \\
& = \frac{E_{\mathcal{A}}[E(Y_j, \bar{S}_j=1\mid \bar{z}_j, c(\bar{z}_j), \bar{\gamma_j},\mathcal{A})]}{E_{\mathcal{A}}[\Pr(\bar{S}_j=1\mid \bar{z}_j,c(\bar{z}_j), \bar{\gamma_j},\nu_j,\mathcal{A})]}.  \tag{A.2}
\end{align}
$E_{\mathcal{A}}[E(Y_j, \bar{S}_j=1\mid \bar{z}_j, c(\bar{z}_j), \bar{\gamma_j},\mathcal{A})]$ is obtained by marginalizing over the  distributions of the set of temporally preceding variables $\mathcal{A}=(\bar{y}_{j-1}, \bar{w}_{j}, \bar{r}_{j}, x_{0} )$. That is, 
\begin{align*}
E&[Y_j(\bar{z}_j), \bar{S}_j(\bar{z}_j)=1] \nonumber \\
& = E[Y_j(\bar{z}_j), \bar{S}_{j-1}(\bar{z}_{j-1})=1 \mid S_{j}(\bar{z}_{j-1})=1]\Pr[S_{j}(\bar{z}_{j-1})=1]  \\
& = \sum_{w_j} E[Y_j(\bar{z}_j), \bar{S}_{j-1}(\bar{z}_{j-1})=1 \mid S_{j}(\bar{z}_{j})=1, w_j]  \\
& \quad \times \Pr[S_{j}(\bar{z}_{j})=1 \mid w_j]p(w_j)  \\
& = \sum_{r_j}\sum_{w_j} E[Y_j(\bar{z}_j), \bar{S}_{j-1}(\bar{z}_{j-1})=1 \mid S_{j}(\bar{z}_{j})=1, w_j,r_j]  \\
& \quad \times \Pr[S_{j}(\bar{z}_{j})=1 \mid w_j,r_j]p(w_j\mid r_j)p(r_j)  \\
& = \int_{y_{j-1}}\sum_{r_j}\sum_{w_j} E[Y_j(\bar{z}_j), \bar{S}_{j-1}(\bar{z}_{j-1})=1 \mid S_{j}(\bar{z}_{j})=1, w_j,r_j,y_{j-1}]  \\
& \quad \times \Pr[S_{j}(\bar{z}_{j})=1 \mid w_j,r_j,y_{j-1}]p(w_j\mid r_j,y_{j-1})p(r_j\mid y_{j-1}) p(y_{j-1})  \\
& \vdots  \\
& = \sum_{x_0}\int_{\bar{y}_{j-1}}\sum_{\bar{r}_j}\sum_{\bar{w}_j} E[Y_j(\bar{z}_j)\mid \bar{y}_{j-1},\bar{w}_j,\bar{r}_j,\bar{S}_{j}(\bar{z}_{j})=1, x_0]  \\
& \quad \times \prod^j_{k=0} \Bigl( \Pr[S_{k}(\bar{z}_{k})=1 \mid \bar{y}_{k-1},\bar{w}_k,\bar{r}_k,\bar{S}_{k-1}(\bar{z}_{k-1})=1, x_0]  \\
& \quad \times 
p(w_k\mid \bar{y}_{k-1},\bar{w}_{k-1},\bar{r}_k,\bar{S}_{k-1}(\bar{z}_{k-1})=1,x_0)  \\
& \quad \times p(r_k\mid \bar{y}_{k-1},\bar{w}_{k-1},\bar{r}_{k-1},\bar{S}_{k-1}(\bar{z}_{k-1})=1,x_0)  \\
& \quad \times p(y_{k-1}\mid \bar{y}_{k-2},\bar{w}_{k-1},\bar{r}_{k-1},\bar{S}_{k-1}(\bar{z}_{k-1})=1,x_0) \Bigr) \\
& \quad \times p(x_{0}) d\bar{y}_{j-1} \quad (Assumption \quad 1) \\
& = \sum_{x_0}\int_{\bar{y}_{j-1}}\sum_{\bar{r}_j}\sum_{\bar{w}_j} E[Y_j \mid \bar{y}_{j-1},\bar{w}_j,\bar{z}_{j},\bar{r}_j,\bar{s}_{j}=1, x_0]  \\
& \quad \times \prod^j_{k=0}\Bigl( \Pr[S_{k}=1 \mid \bar{y}_{k-1}, \bar{z}_{k}, \bar{w}_k,\bar{r}_k,\bar{s}_{k-1}=1, x_0]  \\
& \quad \times 
p(w_k\mid \bar{y}_{k-1},\bar{z}_{k-1},\bar{w}_{k-1},\bar{r}_k,\bar{s}_{k-1}=1, x_0)  \\
& \quad \times p(r_k\mid \bar{y}_{k-1},\bar{z}_{k-1}, \bar{w}_{k-1},\bar{r}_{k-1},\bar{s}_{k-1}=1,x_0)  \\
& \quad \times p(y_{k-1}\mid \bar{y}_{k-2},\bar{z}_{k-1},\bar{w}_{k-1},\bar{r}_{k-1},\bar{s}_{k-1}=1,x_0) \Bigr) \\
& \quad \times p(x_{0}) d\bar{y}_{j-1}  \\
& = E_{\mathcal{A}}[E(Y_j, \bar{S}_j=1\mid \bar{z}_j, c(\bar{z}_j), \bar{\gamma_j},\mathcal{A})]. 
\end{align*}
The expectation $E[Y_j \mid \bar{y}_{j-1}, \bar{z}_{j}, \bar{w}_{j}, \bar{r}_{j}, \bar{s}_{j}=1, x_{0}]$ is identified up to $\gamma_j$ and $c(\bar{z}_j)$ by Assumptions 4 and 5. In particular,
\begin{align*}
E&[Y_j \mid \bar{y}_{j-1}, \bar{z}_{j}, \bar{w}_{j}, \bar{r}_{j}, \bar{s}_{j}=1, x_{0}] \nonumber \\ 
&=  E[Y_j\mid \bar{y}_{j-1}, \bar{z}_{j}, \bar{w}_{j}, \bar{r}_{j}=1, \bar{s}_{j}=1, x_{0}] + I_{(\bar{r}_{j}=\{1,\ldots,1,0\})} \times \gamma_j  \nonumber \\
& \quad - c(\bar{z}_j)\times (1 - \Pr[z_j \mid \bar{y}_{j-1}, \bar{z}_{j-1}, \bar{w}_{j}, \bar{r}_{j}=1, \bar{s}_{j}=1, x_{0})). 
\end{align*}
Note that, by Assumption 4 we have that 
\begin{align*}
p&(y_{k-1} \mid \bar{y}_{k-2}, \bar{z}_{k-1},\bar{w}_{k-1}, \bar{r}_{k-2}=1,r_{k-1}=0, \bar{s}_{k-1}=1,x_0) \\
& = p(y_{k-1} + \gamma_{k-1} \mid \bar{y}_{k-2}, \bar{z}_{k-1},\bar{w}_{k-1}, \bar{r}_{k-1}=1, \bar{s}_{k-1}=1,x_0), 
\end{align*}
and for $t<k-1$ we have that
\begin{align*}
p&(y_{k-1} \mid \bar{y}_{k-2}, \bar{z}_{k-1},\bar{w}_{k-1}, \bar{r}_{t-1}=1, \{r_t=0,\ldots,r_{k-1}=0 \}, \bar{s}_{k-1}=1,x_0)\\
&  =  p(y_{k-1} \mid \bar{y}_{k-2}, \bar{z}_{k-1},\bar{w}_{k-1}, \bar{r}_{k-1}=1, \bar{s}_{k-1}=1, x_0).
\end{align*}
By Assumption 4, for the time-varying confounder we have that 
\begin{align*}
p&(w_k \mid \bar{y}_{k-1}, \bar{z}_{k-1}, \bar{w}_{k-1}, \bar{r}_{t-1}=1, \{r_t=0,\ldots,r_{k-1}=0 \}, \bar{s}_{k}=1, x_0) \\
& = p(w_k \mid \bar{y}_{k-1}, \bar{z}_{k-1}, \bar{w}_{k-1}, \bar{r}_{k}=1, \bar{s}_{k}=1, x_0) 
\end{align*}
for $t\leq k-1$.

For identification of the denominator in Eq A.2 we have that $E_{\mathcal{A}}[\Pr(\bar{S}_j=1\mid \bar{z}_j,c(\bar{z}_j), \bar{\gamma_j},\nu_j,\mathcal{A})]$ is obtained by marginalizing over the  distributions of the set of temporally preceding variables $\mathcal{A}=(\bar{y}_{j-1}, \bar{w}_{j}, \bar{r}_{j}, x_{0} )$. That is, 
\begin{align*}
\Pr&[\bar{S}_j(\bar{z}_j)=1] \\ 
& = \sum_{x_0}\int_{\bar{y}_{j-1}}\sum_{\bar{r}_j}\sum_{\bar{w}_j} \prod^j_{k=0} \Bigl( \Pr[S_{k}(\bar{z}_{k})=1 \mid \bar{y}_{k-1},\bar{w}_k,\bar{r}_k,\bar{S}_{k-1}(\bar{z}_{k-1})=1, x_0]  \\
& \quad \times 
p(w_k\mid \bar{y}_{k-1},\bar{w}_{k-1},\bar{r}_k,\bar{S}_{k-1}(\bar{z}_{k-1})=1,x_0)  \\
& \quad \times p(r_k\mid \bar{y}_{k-1},\bar{w}_{k-1},\bar{r}_{k-1},\bar{S}_{k-1}(\bar{z}_{k-1})=1,x_0)  \\
& \quad \times p(y_{k-1}\mid \bar{y}_{k-2},\bar{w}_{k-1},\bar{r}_{k-1},\bar{S}_{k-1}(\bar{z}_{k-1})=1,x_0) \Bigr) \\
& \quad \times p(x_{0}) d\bar{y}_{j-1} \quad (Assumption \quad 1)  \\
&= \sum_{x_0}\int_{\bar{y}_{j-1}}\sum_{\bar{r}_j}\sum_{\bar{w}_j} \prod^j_{k=0}\Bigl( \Pr[S_{k}=1 \mid \bar{y}_{k-1}, \bar{z}_{k},\bar{w}_k,\bar{r}_k,\bar{s}_{k-1}=1, x_0] \\
& \quad \times 
p(w_k\mid \bar{y}_{k-1},\bar{z}_{k-1},\bar{w}_{k-1},\bar{r}_k,\bar{s}_{k-1}=1,x_0) \\
& \quad \times p(r_k\mid \bar{y}_{k-1},\bar{z}_{k-1},\bar{w}_{k-1},\bar{r}_{k-1},\bar{s}_{k-1}=1,x_0) \\
& \quad \times p(y_{k-1}\mid \bar{y}_{k-2},\bar{z}_{k-1},\bar{w}_{k-1},\bar{r}_{k-1},\bar{s}_{k-1}=1,x_0) \Bigr) \\
& \quad \times p(x_{0}) d\bar{y}_{j-1} \\
& =E_{\mathcal{A}}[\Pr(\bar{S}_j=1\mid \bar{z}_j,c(\bar{z}_j), \bar{\gamma_j},\nu_j,\mathcal{A})]. 
\end{align*}
For identification of $\Pr[S_{k}=1 \mid \bar{y}_{k-1}, \bar{z}_{k}, \bar{w}_{k}, \bar{r}_{k}, \bar{s}_{k}=1, x_{0}]$ we first note that, for all individuals who participates at the $k$th wave
\begin{align*}
\Pr[S_k=1 \mid \bar{y}_{k-1}, \bar{z}_{k},\bar{w}_k,\bar{r}_k=1,\bar{s}_{k-1}=1, x_0]=1.
\end{align*}
For those individuals who have dropped out, $r_k=0$, we have that 
\begin{align}\label{PrSdrop}
\Pr&[S_k=1 \mid \bar{y}_{k-1},\bar{z}_{k},\bar{w}_k, r_k=0,\bar{s}_{k-1}=1, x_0] \nonumber \\ 
& = \frac{\Pr[z_k \mid \bar{y}_{k-1},\bar{z}_{k-1},\bar{w}_k,r_k=0,\bar{s}_k = 1, x_0]}{\Pr[z_k \mid \bar{y}_{k-1},\bar{z}_{k-1},\bar{w}_k,r_k=0,\bar{s}_{k-1}= 1, x_0]} \nonumber \\
& \quad \times \Pr[S_k = 1 \mid \bar{y}_{k-1},\bar{z}_{k-1},\bar{w}_k,r_k=0,\bar{s}_{k-1}= 1, x_0],  \tag{A.3} 
\end{align} 
where the numerator on the rhs is identified by Assumption 5 and the denominator is a function of $\nu_j$ in Assumption 8. That is,
\begin{align*}
&\Pr[z_k \mid \bar{y}_{k-1},\bar{z}_{k-1},\bar{w}_k,r_k=0,\bar{s}_{k-1}= 1, x_0] \nonumber \\
& = \nu_j \Pr[S_k=0 \mid \bar{y}_{k-1},\bar{z}_{k-1},\bar{w}_k,r_k=0,\bar{s}_{k-1}= 1, x_0]\nonumber \\
& \quad + \Pr[z_k \mid \bar{y}_{k-1}, \bar{z}_{k-1},\bar{w}_k,r_k=0, \bar{s}_{k}=1, x_0] \nonumber \\
& \quad \times \Pr[S_{k}=1 \mid \bar{y}_{k-1},\bar{z}_{k-1},\bar{w}_k,r_k=0,\bar{s}_{k-1}= 1, x_0].
\end{align*}
We further have that the last expression in Eq A.3, 
\begin{align*}
&\Pr[S_{k}=1 \mid \bar{y}_{k-1},\bar{z}_{k-1},\bar{w}_k,r_k=0,\bar{s}_{k-1}= 1, x_0]  \\
& = \Pr[w_k \mid \bar{y}_{k-1},\bar{z}_{k-1},\bar{w}_{k-1},r_k=0,\bar{s}_{k}= 1, x_0]  \\
& \quad \times \Pr[R_k=0 \mid \bar{y}_{k-1},\bar{z}_{k-1},\bar{w}_{k-1},\bar{r}_{k-1},\bar{s}_{k}= 1, x_0]  \\
& \quad \times \Pr[S_{k}=1 \mid \bar{y}_{k-1},\bar{z}_{k-1},\bar{w}_{k-1},\bar{r}_{k-1},\bar{s}_{k-1}= 1, x_0]  \\
& \quad \times \frac{1}{\Pr[w_k \mid \bar{y}_{k-1},\bar{z}_{k-1},\bar{w}_{k-1},r_k=0,\bar{s}_{k-1}= 1, x_0]}  \\
& \quad \times \frac{1}{\Pr[R_k=0 \mid \bar{y}_{k-1},\bar{z}_{k-1},\bar{w}_{k-1},\bar{r}_{k-1},\bar{s}_{k-1}= 1, x_0]} \quad (Assumption \quad 8) \\
& = \Pr[R_k=0 \mid \bar{y}_{k-1},\bar{z}_{k-1},\bar{w}_{k-1},\bar{r}_{k-1},\bar{s}_{k}= 1, x_0]  \\
& \quad \times \Pr[S_{k}=1 \mid \bar{y}_{k-1},\bar{z}_{k-1},\bar{w}_{k-1},\bar{r}_{k-1},\bar{s}_{k-1}= 1, x_0]  \\
& \quad \times \frac{1}{\Pr[R_k=0 \mid \bar{y}_{k-1},\bar{z}_{k-1},\bar{w}_{k-1},\bar{r}_{k-1},\bar{s}_{k-1}= 1, x_0]}, \\
\end{align*}
where
\begin{align*}
& \Pr[R_k=0 \mid \bar{y}_{k-1}, \bar{z}_{k-1},\bar{w}_{k-1},\bar{r}_{k-1},\bar{s}_{k-1}= 1, x_0]  \\
& = \Pr[R_k=0 \mid \bar{y}_{k-1},\bar{z}_{k-1},\bar{w}_{k-1},\bar{r}_{k-1},\bar{s}_{k}= 1, x_0]   \\
& \quad \times \Pr[S_k=1 \mid \bar{y}_{k-1}, \bar{z}_{k-1},\bar{w}_{k-1},\bar{r}_{k-1},\bar{s}_{k-1}= 1, x_0] \\
& \quad + \Pr[S_k=0 \mid \bar{y}_{k-1}, \bar{z}_{k-1},\bar{w}_{k-1},\bar{r}_{k-1},\bar{s}_{k-1}= 1, x_0],
\end{align*}
by using Assumption 5 and 8.

For identification of the last expression in Eq A.1 we first note that $\Pr[\bar{S}_j(\bar{z}'_j)\neq 1 \mid \bar{S}_j(\bar{z}_j)=1]= 0$, by Assumption 6. For the contrasting regime $\bar{z}'_j$, we have that
\begin{align*}
&\Pr[\bar{S}_j(\bar{z}_j)\neq 1 \mid \bar{S}_j(\bar{z}'_j)=1] \nonumber \\
&= 1 - \Pr[\bar{S}_j(\bar{z}_j)= 1 \mid \bar{S}_j(\bar{z}'_j)=1] \nonumber \\
&= 1- \frac{\Pr[\bar{S}_j(\bar{z}_j)=\bar{S}_j(\bar{z}'_j)= 1)}{\Pr[\bar{S}_j(\bar{z}'_j)= 1]} \nonumber \\
&= 1-\frac{\Pr[\bar{S}_j = 1 \mid \bar{z}_j)}{\Pr[\bar{S}_j = 1 \mid \bar{z}'_j]} 
\end{align*}
by Assumptions 1 and 6. Because

\begin{align*}
\Pr&[\bar{S}_j(\bar{z}_j)=\bar{S}_j(\bar{z}'_j)=1] \\ 
& = \sum_{x_0}\int_{\bar{y}_{j-1}}\sum_{\bar{r}_j}\sum_{\bar{w}_j} \prod^j_{k=0} \\
& \quad \Bigl(\Pr[S_{k}(\bar{z}_{k})=S_k(\bar{z}'_j)=1 \mid \bar{y}_{k-1},\bar{w}_k,\bar{r}_k,\bar{S}_{k-1}(\bar{z}_{k-1})=\bar{S}_{k-1}(\bar{z}'_{k-1})=1, x_0] \\
& \quad \times 
p(w_k\mid \bar{y}_{k-1},\bar{w}_{k-1},\bar{r}_k,\bar{S}_{k-1}(\bar{z}_{k-1})=\bar{S}_{k-1}(\bar{z}'_{k-1})=1,x_0)  \\
& \quad \times p(r_k\mid \bar{y}_{k-1},\bar{w}_{k-1},\bar{S}_{k-1}(\bar{z}_{k-1})=\bar{S}_{k-1}(\bar{z}'_{k-1})=1,x_0)  \\
& \quad \times p(y_{k-1}\mid \bar{y}_{k-2},\bar{w}_{k-1},\bar{S}_{k-1}(\bar{z}_{k-1})=\bar{S}_{k-1}(\bar{z}'_{k-1})=1,\bar{r}_{k-1},x_0) \Bigr) \\
& \quad \times p(x_{0}) d\bar{y}_{j-1} \quad  (Assumption \quad 6)  \\
& = \sum_{x_0}\int_{\bar{y}_{j-1}}\sum_{\bar{r}_j}\sum_{\bar{w}_j} \prod^j_{k=0} \\
& \quad \Bigl(\Pr[S_{k}(\bar{z}_{k})=1 \mid \bar{y}_{k-1},\bar{w}_k,\bar{r}_k,\bar{S}_{k-1}(\bar{z}_{k-1})=1, x_0] \\
& \quad \times 
p(w_k\mid \bar{y}_{k-1},\bar{w}_{k-1},\bar{r}_k,\bar{S}_{k-1}(\bar{z}_{k-1})=1,x_0)  \\
& \quad \times p(r_k\mid \bar{y}_{k-1},\bar{w}_{k-1},\bar{r}_{k-1},\bar{S}_{k-1}(\bar{z}_{k-1})=1,x_0)  \\
& \quad \times p(y_{k-1}\mid \bar{y}_{k-2},\bar{w}_{k-1},\bar{r}_{k-1},\bar{S}_{k-1}(\bar{z}_{k-1})=1,x_0) \Bigr) \\
& \quad \times p(x_{0}) d\bar{y}_{j-1} \quad  (Assumption \quad 1)  \\
&= \sum_{x_0}\int_{\bar{y}_{j-1}}\sum_{\bar{r}_j}\sum_{\bar{w}_j} \prod^j_{k=0} \Bigl(\Pr[S_{k}=1 \mid \bar{y}_{k-1}, \bar{z}_{k},\bar{w}_k,\bar{r}_k,\bar{s}_{k-1}=1, x_0] \\
& \quad \times 
p(w_k\mid \bar{y}_{k-1},\bar{z}_{k-1},\bar{w}_{k-1},\bar{r}_k,\bar{s}_{k-1}=1,x_0) \\
& \quad \times p(r_k\mid \bar{y}_{k-1},\bar{z}_{k-1},\bar{w}_{k-1},\bar{s}_{k-1}=1,x_0) \\
& \quad \times p(y_{k-1}\mid \bar{y}_{k-2},\bar{z}_{k-1},\bar{w}_{k-1},\bar{r}_{k-1},\bar{s}_{k-1}=1,x_0)\Bigr) \\
& \quad \times p(x_{0}) d\bar{y}_{j-1} \\
& =E_{\mathcal{A}}[\Pr(\bar{S}_j=1\mid \bar{z}_j,c(\bar{z}_j), \bar{\gamma_j},\nu_j,\mathcal{A})]. 
\end{align*}

\section*{A.4: Details on algorithm for estimation of $\tau$ in Section 5.3.}
Here we present an algorithm for estimation of $\tau$ in [2] using the G-computation formula. The general approach is to specify models for the observed data as we did in Section 5 and then to use assumptions in Section 3 with embedded sensitivity parameters to identify the causal effect estimate as described in Section 4. The algorithm can be summarized in the following five steps: 
\begin{enumerate}
\item Sample the observed data posteriors as described in Section 5.
\item MC-sampling. For each posterior sample of the parameters sample pseudo data $(\bar{y}^*_{j-1},\bar{w}^*_{j},\bar{r}_{j}^*,\bar{s}_{j}^*,x_0^*)$ of size $N^*$. Specifically, for a fixed regime $\bar{z}_{j-1}$, sequentially compute conditional expectations using pseudo data, the sensitivity parameters and the posterior sample of the model parameters. Further sample new data from this distribution. Note that $s_{k}^*$ and $y^*_{k-1}$ is sampled for the subset that satisfies $\bar{s}_{k-1}^*=1$, while $r_{k}^*$ and $w^*_{k}$ is sampled for the subset that satisfies $\bar{s}_{k}^*=1$. Note that, $s_0^*=1$ and $r_0^*=1$ for all individuals, and that $s_k^*=0$ if $s_{k-1}^*=0$ and $r_k^*=0$ if $r_{k-1}^*=0$. Sample one set from the prior distribution of the sensitivity parameters. 
\item Implement G-computation for $\bar{z}_j$, and similarly for $\bar{z}'_j$, using the pseudo data and sensitivity parameters from Step 2 by estimating the parameters of interest:
\begin{enumerate}
\item Compute $E[Y_j \mid \bar{y}_{j-1}, \bar{z}_j, \bar{w}_{j},\bar{r}_{j}, \bar{s}_j=1, x_0]$ 
for a fixed regime $\bar{z}_j$, denoted by $\phi_j(\bar{z}_j)$ as follows:
\begin{align*}
\phi_j(\bar{z}_j)=&\mu_j(\bar{y}^*_{j-1}, \bar{z}_j, \bar{w}^*_{j}, x^*_0) + I_{(r^*_j=0)} \times \gamma_j \\
&- c(\bar{z}_j)\times (1-\pi_{z_j}(\bar{y}^*_{j-1}, \bar{z}_{j-1}, \bar{w}^*_{j}, x^*_0)).
\end{align*}
\item Compute $\prod^j_{k=0} \Pr[S_{k}=1 \mid \bar{z}_{k}, \bar{w}_k,\bar{r}_k,\bar{y}_{k-1},\bar{S}_{k-1}=1, x_0]$ 
, denoted by $\chi_j(z_j)$. 
For the subset that satisfies $r_j^*=1$, 
$$\chi_k(z_k)=1,$$
otherwise, when $r_j^*=0$,
\begin{align*}
\chi_j(z_j)=\prod^j_{k=0} \frac{\pi_{z_{k}}(\bar{y}^*_{k-1}, \bar{z}_{k-1}, \bar{w}^*_{k}, x_0^*)A_k}{B_k}
\end{align*}
where 
\begin{align*}
&A_k=(1 - \pi_{R_k}(\bar{y}^*_{k-1}, \bar{z}_{k-1}, \bar{w}^*_{k-1}, x_0^*))\pi_{S_k}(\bar{y}^*_{k-1}, \bar{z}_{k-1}, \bar{w}^*_{k-1}, x_0^*)\\
&\qquad \times 1/[(1 - \pi_{R_k}(\bar{y}^*_{k-1}, \bar{z}_{k-1}, \bar{w}^*_{k-1}, x_0^*))\pi_{S_k}(\bar{y}^*_{k-1}, \bar{z}_{k-1}, \bar{w}^*_{k-1}, x_0^*) \\
&\qquad +(1-\pi_{S_k}(\bar{y}^*_{k-1}, \bar{z}_{k-1}, \bar{w}^*_{k-1}, x_0^*))],
\end{align*}
and
\begin{align*}
&B_k=\pi_{z_{k}}(\bar{y}^*_{k-1}, \bar{z}_{k-1}, \bar{w}^*_{k}, x_0^*)(\nu_{z_k} - \nu_{z_k} A_k + A_k).
\end{align*}
\item Implement Monte Carlo integration using the pseudo data and sensitivity parameters to compute $\Pr[\bar{S}_{j}=1 \mid \bar{z}_j]$ 
, denoted by $\hat{p}_{s_j \mid \bar{z}_j}$: 
\begin{align*}
\hat{p}_{s_j \mid \bar{z}_j}=\frac{\sum \chi_j(\bar{z}_j)}{N^*}.
\end{align*}
\item Implement Monte Carlo integration using the pseudo data and sensitivity parameters to compute $E[Y_j, \bar{S}_j=1 \mid \bar{z}_j]$ 
, denoted by $\hat{\mu}_{y_j, \bar{s}_j=1 \mid \bar{z}_j}$: 
\begin{align*}
\hat{\mu}_{y_j, \bar{s}_j=1 \mid \bar{z}_j}= \frac{\sum \phi_j(\bar{z}_j)\chi_j(\bar{z}_j)}{N^*}.
\end{align*}
\end{enumerate}
\item Use the quantities in step (a)-(d) above to compute one posterior sample of $\tau$ as defined in [4]-[5]:
\begin{align*}
\sum_{j=1}^J \frac{\hat{p}_{s_j \mid \bar{z}_j}}{\sum_{k=1}^J \hat{p}_{s_k \mid \bar{z}_k}} \Bigg\{ \frac{\hat{\mu}_{y_j, \bar{s}_j=1 \mid \bar{z}_j}}{\hat{p}_{s_j \mid \bar{z}_j}} - \frac{\hat{\mu}_{y_j, \bar{s}_j=1 \mid \bar{z}'_j}}{\hat{p}_{s_j \mid \bar{z}'_j}}
- \Delta_{j} \left(1- \frac{\hat{p}_{s_j\mid \bar{z}_j}}{\hat{p}_{s_j\mid \bar{z}'_j}}\right)\Bigg\}.
\end{align*}
\item Repeat step 2 - 4 for each of the posterior sample of the parameters. 
\end{enumerate}

\section*{A.5: Details on the simulation study in Section 6.}
For simplicity of comparison we estimate $\mathrm{E}[Y_j(\bar{z}_j)-Y_j(\bar{z}'_j)\mid \bar{s}_j=1]$ and set $\Delta_{\bar{z}'_{j}}=0$, $\gamma_j=0$, and $c(z_j)=c(z'_j)=0$. 
The causal contrasts are thus estimated by computing
$\frac{\sum \mu_j(\bar{y}^*_{j-1}, \bar{z}_j, \bar{w}^*_{j}, x^*_0)}{N^*} - \frac{\sum \mu_j(\bar{y}^*_{j-1}, \bar{z}'_j, \bar{w}^*_{j}, x^*_0)}{N^*},$ and the weights in [4] are estimated by computing 
$\frac{\frac{\sum_{1}^{N^*} \pi_{S_j}(\bar{y}^*_{j-1}, \bar{z}_{j-1}^*, \bar{w}^*_{j-1}, x_0^*)}{N^*}}{\sum_{1}^J \frac{\sum_{1}^{N^*} \pi_{S_k}(\bar{y}^*_{k-1}, \bar{z}_{k-1}^*, \bar{w}^*_{k-1}, x_0^*)}{N^*}}.$ For IPTW-W, IPTW-SW, and TMLE, the causal effect was obtained by pooling the causal contrasts  using the following weights $\frac{\Pr[\bar{S}_j=1]}{\sum_{k=1}^J \Pr[\bar{S}_k=1]}$.

Data was generated in R as follows:
\begin{lstlisting}
#######################
## Data generation
n.sim <- 1000

## j=0
X0 <- runif(n.sim)
W0 <- rbinom(n.sim, 1, expit(-2 + 0.25*X0 - 2.5*X0^2 + 5*X0^3))
Y0 <- rnorm(n.sim, 0.5 - 0.1*W0 - 0.1*X0 + 0.25*X0^2 - 0.5*X0^3, 
	    sd=0.1)
## j=1
W1 <- rbinom(n.sim, 1, expit(-2 + 0.25*X0 - 2.5*X0^2 + 5*X0^3 + 
	    0.25*W0))
Z1 <- rbinom(n.sim, 1, expit(-5 + 0.3*W0 + 0.6* W1 + X0 - 4*X0^2 + 
	    6*X0^3 + 0.5*W1*X0 - W1*X0^2 +  2*W1*X0^3))
Y1 <- rnorm(n.sim, 0.5 - 0.05*Z1 - 0.1*W1 + 0.25*Y0 - 0.1*X0 + 
	    0.25*X0^2 - 0.25*X0^3, sd=0.1)
## j=2
W2 <- rbinom(n.sim, 1, expit(-2 + 0.25*X0 - 2.5*X0^2 + 5*X0^3 + 
	    0.25*W1))
Z2 <- rbinom(n.sim, 1, expit(-5 + 0.3*W1 + 0.6*W2 + X0 - 4*X0^2 + 
	    8*X0^3 + 0.5*W2*X0 - W2*X0^2 +  2*W2*X0^3))
Y2 <- rnorm(n.sim, 0.5 - 0.05*Z2 - 0.1*W2 + 0.25*Y1 -  0.1*X0 + 
	    0.25*X0^2 - 0.25*X0^3, sd=0.1)

SimData <- data.frame(X0, W0, Y0, W1, Z1, Y1, W2, Z2, Y2)
\end{lstlisting}